\documentclass[aps,prb,reprint,amsmath,amssymb]{revtex4-2}
\usepackage{graphicx}
\usepackage{bm}
\usepackage{amsthm,amssymb,amsmath,amsfonts,latexsym,enumerate,float}
\usepackage[utf8]{inputenc}
\usepackage{cmap}
\allowdisplaybreaks

\begin{document}


\title{Features of magnetization and spin reorientation\\ in weak ferrimagnets of the YF\lowercase{e}$_{1-x}$C\lowercase{r}$_x$O$_3$ type}


\author{Alexander Moskvin}
\email{alexander.moskvin@urfu.ru}
\affiliation{Ural Federal University, 620083 Ekaterinburg, Russia}
\affiliation{M.N. Mikheev lnstitute of Metal Physics of Ural Branch of Russian Academy of Sciences, 620108 Ekaterinburg, Russia}
\author{Evgenii Vasinovich}
\email{evgeny.vasinovich@urfu.ru}
\affiliation{Ural Federal University, 620083 Ekaterinburg, Russia}


\begin{abstract}
A brief critical review is given of the 50-year history of experimental and theoretical studies of the magnetic properties of a new promising class of weak ferrimagnets such as RFe$_{1-x}$Cr$_x$O$_3$ with a non-magnetic R-ion (R\,=\,La , Y, Lu), i.e. systems with competing signs of the Dzyaloshinsky vectors Fe--Fe, Cr--Cr and Fe--Cr. The spin Hamiltonian of the system is considered taking into account isotropic exchange, antisymmetric Dzyaloshinsky-Moriya exchange, and second- and fourth-order single-ion anisotropy. Within the framework of the molecular field approximation, calculations were made of the Neel temperatures, the average magnetic moments of $3d$ ions, total and partial magnetizations, and effective anisotropy constants. The existence in the model system YFe$_{1-x}$Cr$_x$O$_3$ of two regions of negative magnetization $0.25 \leq x\leq 0.5$ and $x\approx 0.8$ with the corresponding magnetization reversal points reaching room temperature at $x\approx 0.45$. The phenomenon of spin reorientation observed for single-crystal samples in a wide range of concentrations is explained by a sharp decrease in the contribution of antisymmetric exchange to magnetic anisotropy with increasing deviation from the parent compositions and competition between the contributions of single-ion anisotropy of Fe and Cr ions. It has been suggested that the spatial orientation of the Neel ${\bf G}$ vector and the $G_{xyz}$ configuration are the reason for the small value of saturation magnetization observed experimentally for compositions inside or near the region of negative magnetization.

\end{abstract}

\maketitle


\section{Introduction}

Rare earth orthoferrites RFeO$_3$ and orthochromites RCrO$3$ (R = Y, or rare earth ion) have attracted and continue to attract special attention of researchers for many decades due to the combination of unique magnetic, magnetooptical, magnetoelastic and magnetic resonance properties, primarily weak ferro- and antiferromagnetism, spin-reorientation transitions, negative magnetization phenomena (magnetization reversal), magnetoelectric effects. Orthoferrites and orthochromites remain a relevant object not only of fundamental theoretical and experimental, but also of applied research related to the perspective for using these materials in various innovative spintronics devices.
 
The features of $4f$--$3d$ interaction and, first of all, antisymmetric Dzyaloshinsky-Moriya exchange remain at the center of fundamental research on orthoferrites and orthochromites. The development of the microscopic theory of the Dzyaloshinsky-Moriya interaction\,\cite{1977,thesis,JMMM-2016,CM-2019,JETP-2021}, in particular the orientation and sign of the Dzyaloshinsky vector for pairs of $3d$ ions, made it possible not only to give a quantitative description weak ferro- and antiferromagnetism of orthoferrites and orthochromites, but also to predict a new class of magnets with competing signs of the Dzyaloshinsky vector based on mixed weak ferromagnets RFe$_{1-x}$Cr$_x$O$_3$\,\cite{WFIM-1,WFIM-2,YLuFeCrO3,JETP-1983,APP-1985,NdFeCrO3}, Fe$_{1-x}$Cr$_x$BO$_3$\,\cite{FeCrBO3}, Mn$_{1- x}$Ni$_x$CO$_3$\,\cite{MnNiCO3,FeNi}, Fe$_{2-x}$Cr$_x$O$_3$\,\cite{FeCr2O3}, named {\it weak ferrimagnets}.

The first experimental studies of mixed orthoferrites-orthochromites YFe$_{1-x}$Cr$_x$O$_3$ were carried out at Moscow University by A.~M. Kadomtseva and co-workers more than 50 years ago on both polycrystalline and single-crystalline samples\,\cite{WFIM-1972,WFIM-1972a,WFIM-1976,WFIM-1,WFIM-2,KP}. These studies were initiated, in particular, by the predictions of Goodenough and Kanamori\,\cite{Good}, according to which the superexchange integral Fe$^{3+}$--O$^{2-}$--Cr$^{3+ }$ should have a ``ferromagnetic sign'', which could lead to a fundamental restructuring of the magnetic structure of mixed compositions. However, both experimental and theoretical studies have shown that in these systems we are not dealing with the competition of isotropic exchange interactions, but with the peculiarities of the Dzyaloshinsky-Moriya interaction for various $3d$ ions.

Magnetic (torsion magnetic balances, vibrating magnetometer, torques) and neutron diffraction measurements have revealed a number of unique magnetic properties of YFe$_{1-x}$Cr$_x$O$_3$, such as a sharp drop of one or two orders of magnitude in spontaneous magnetization in substituted compositions, magnetization reversal, the appearance of spin reorientation, which is absent in the ``parent'' compositions YFeO$_3$ and YCrO$_3$. These features received a natural explanation within the framework of theoretical concepts about the competition of Dzyaloshinsky-Moriya (DM) interactions Fe--Fe, Cr--Cr and Fe--Cr in orthoferrites-orthochromites of the YFe$_{1-x}$Cr$_x$O$_3$ type, in particular, different signs of the Dzyaloshinsky vectors ${\bf d}_{FeFe}$, ${\bf d}_{CrCr}$ and ${\bf d}_{FeCr}$\,\cite{1977,WFIM-1,thesis}, which led to the antiparallel orientation of the average weak ferromagnetic moments of the Fe and Cr subsystems in a wide range of concentrations, that is, to the {\it weak ferrimagnetic} behavior of YFe$_{1-x}$Cr$_x $O$_3$. Thus, a new phenomenon was experimentally discovered and theoretically explained, i.e. weak ferrimagnetism (WFIM) in mixed $3d$ systems with competing signs of the Dzyaloshinsky vector. Just as different signs of the ordinary exchange integral determine different (ferro-antiferro) magnetic order, different signs of Dzyaloshinsky vectors create the possibility of inhomogeneous (ferro-antiferro) ordering of local weak ferromagnetic moments.

Weak ferrimagnets based on orthoferrites-orthochromites of the RFe$_{1-x}$Cr$_x$O$_3$ type (R = Nd, Gd, Dy, Y, Lu)\,\cite{YLuFeCrO3,JETP-1983,APP-1985,NdFeCrO3}, as well as Fe$_{1-x}$Cr$_x$BO$_3$\,\cite{FeCrBO3}, Mn$_{1-x}$Ni$_x$CO$_3$ \,\cite{MnNiCO3}, Fe$_{2-x}$Cr$_x$O$_3$\,\cite{FeCr2O3} were the subject of intensive fundamental theoretical and experimental research in the late 20th century. A new surge of interest in these systems already in the 21st century\,\cite{Kuznetsov,Dahmani2002,Mao2011,Dasari,Bora,Nair,Pomiro,Billoni,Yin2017,LaFeCrO3,Salazar2022,Yang,Liu} is associated with the opened prospects for the practical use of the magnetization reversal phenomenon and the associated effects of ``negative'' magnetization, exchange bias for the creation of various multifunctional spintronics devices, detection of specific magnetoelectric and magnetocaloric properties\,\cite{Yin2017,multi,nomulti}.
 
However, in many ``new'' works devoted to the study of weak ferrimagnetic orthoferrites-orthochromites, and carried out exclusively on polycrystalline samples, mainly of composition YFe$_{0.5}$Cr$_{0.5}$O$_3$, we are faced with both the ambiguity of experimental data and the ambiguity in their interpretation. In particular, this concerns the choice of the calculation scheme for the molecular field approximation\,\cite{Dasari}, incorrect interpretation of the nature of the magnetization reversal phenomenon, for example, as a result of competition between single-ion spin anisotropy and the Dzyaloshinsky-Moriya interaction\,\cite{Mao2011,Bora,LaFeCrO3, Yang, Liu}, neglect of many interactions that are fundamental for these systems, described in sufficient detail in review articles\,\cite{JMMM-2016,CM-2019,JETP-2021}.
The unusual effect of spin reorientation in weak ferrimagnets with a nonmagnetic R-ion (Y, Lu), discovered back in 1972\,\cite{WFIM-1972a}, has also not yet received an adequate description. Moreover, some authors\,\cite{Dasari} associate spin reorientation in the weak ferrimagnet YFe$_{1-x}$Cr$_x$O$_3$ with the simple effect of changing the canting angle of magnetic sublattices.

These circumstances stimulated this work, in which we give a critical review of the literature data, consider the description of the magnetic structure, in particular the magnetization reversal phenomena and spin reorientation in weak ferrimagnets of the YFe$_{1-x}$Cr$_x$O$_3$ type, within the framework of a more rigorous approach developed earlier for ``parent'' orthoferrites and orthochromites in the works\,\cite{WFIM-1,JMMM-2016,CM-2019,JETP-2021}. Section 2 presents the general form of the Hamiltonian for a weak ferrimagnet, with a brief analysis of the role of the various contributions. Section 3 is devoted to the formulation of the molecular field model (MFA) taking into account the multi-sublattice structure of RFe$_{1-x}$Cr$_x$O$_3$ with a non-magnetic R-ion (La, Y, Lu). Section 4 discusses experimental data on the concentration dependence of the Neel temperature of YFe$_{1-x}$Cr$_x$O$_3$ and the corresponding MFA calculations, and provides a comparison of the experimental and calculated temperature dependence of the average magnetic moment in YFe$_{1- x}$Cr$_x$O$_3$. Section 6 is devoted to the nature of spin reorientation in weak ferrimagnets. Section 7 discusses experimental data on the concentration and temperature dependence of magnetization and ``hidden'' noncollinearity in YFe$_{1-x}$Cr$_x$O$_3$ and the magnetization reversal phenomenon, and compares it with MFA calculations. A brief conclusion is presented in Section 8.

\section{Basic spin interactions and the Hamiltonian of a weak ferrimagnet YF\lowercase{e}$_{1-x}$C\lowercase{r}$_x$O$_3$}

\begin{figure}[h]
\centering
\includegraphics[width=0.48\textwidth]{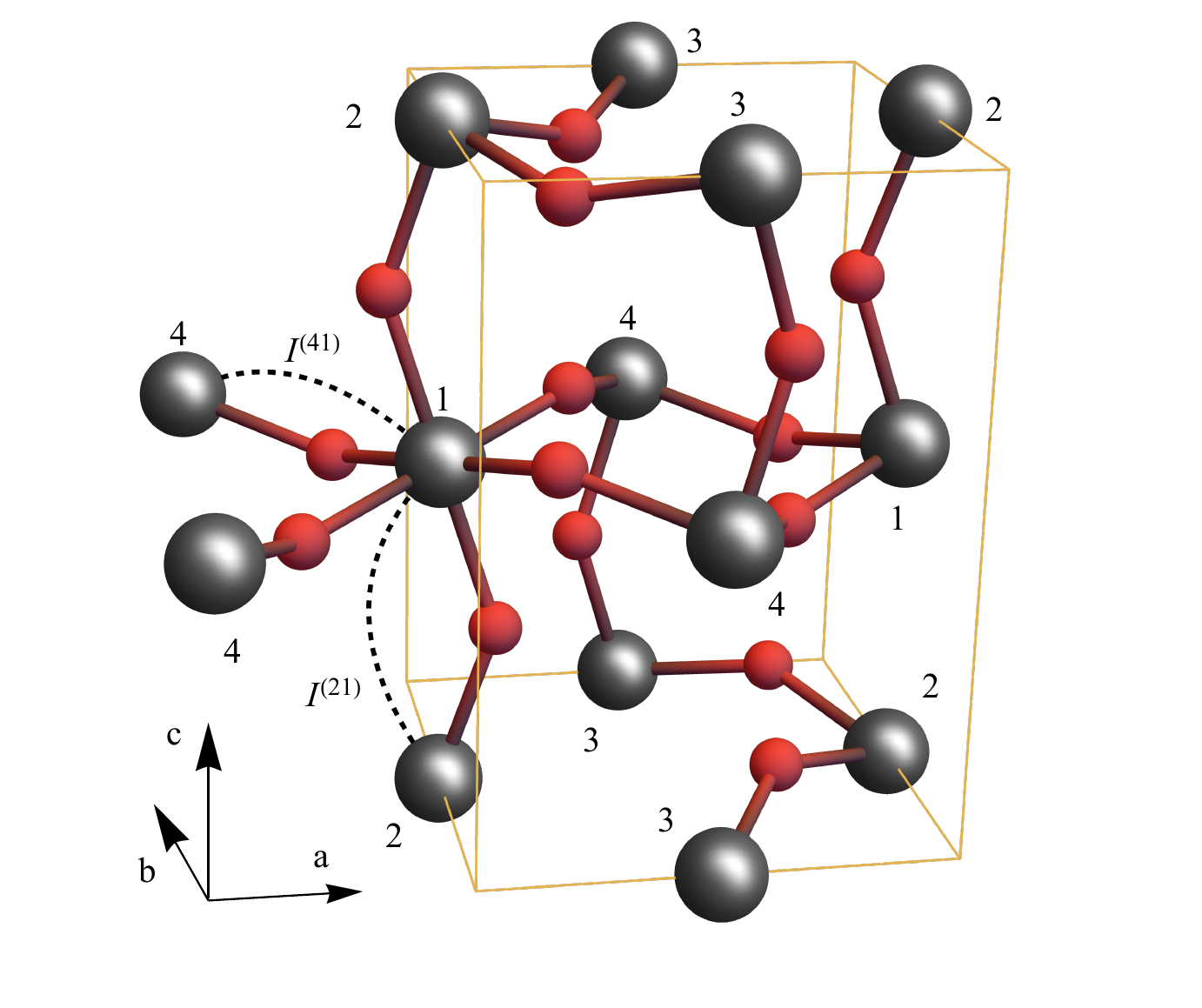}
\caption{
Structure of superexchange bonds; large balls -- Fe$^{3+}$, Cr$^{3+}$ ions, small ones -- O$^{2-}$; 1, 2, 3, 4 -- magnetic ions in four nonequivalent positions
\label{LatticeYFeCrO}}
\end{figure}

Like the ``parent'' orthoferrites YFeO$_3$ and orthochromites YCrO$_3$, mixed systems YFe$_{1-x}$Cr$_x$O$_3$ are orthorhombic perovskites with space group $Pbnm$ ($D_{ 2h}^{16}$)\,\cite{Kuznetsov,Dahmani2002,Mao2011,Dasari,Nair,Pomiro,Billoni,LaFeCrO3,Salazar2022,Yang,Liu}.
As shown in Fig.~\ref{LatticeYFeCrO}, Fe$^{3+}$ (Cr$^{3+}$) ions occupy positions 4b in the unit cell: $1\,(1/2, 0, 0)$, $ 2\,(1/2, 0, 1/2)$, $3\,(0, 1/2, 1/2)$, $4\,(0, 1/2, 0)$. The classical basis vectors of the magnetic structure for the $3d$ sublattice are defined as follows:
\begin{gather}
4S{\bf F}={\bf S}_1+{\bf S}_2+{\bf S}_3+{\bf S}_4\,; \nonumber\\
4S{\bf G}={\bf S}_1-{\bf S}_2+{\bf S}_3-{\bf S}_4\,; \nonumber\\
4S{\bf C}={\bf S}_1+{\bf S}_2-{\bf S}_3-{\bf S}_4\,; \nonumber\\
4S{\bf A}={\bf S}_1-{\bf S}_2-{\bf S}_3+{\bf S}_4\,.
\label{ACFG}
\end{gather}
Here the vector ${\bf G}$ describes the main antiferromagnetic component of the magnetic structure, ${\bf F}$ is the vector of ferromagnetism (``overt canting''  of the sublattices), weak antiferromagnetic components ${\bf C}$ and ${\bf A}$ describe the canting of the magnetic sublattices without the formation of a total magnetic moment (``hidden canting'' of the sublattices). ``Allowed'' spin configurations for the $3d$ sublattice, compatible with the antiferromagnetic sign of the main isotropic superexchange, are denoted as $\Gamma_1\,(A_x, G_y, C_z)$, $\Gamma_2\,(F_x, C_y, G_z) $, $\Gamma_4\,(G_x, A_y, F_z)$, where the only non-zero components of the basis vectors appear in parentheses. Note that in the literature one can find different options for numbering the positions of Fe$^{3+}$ ions (see, for example, the works\,\cite{muon,Amelin}), so that the basis vectors ${\bf G}$, ${\bf C}$, ${\bf A}$ may differ in sign.

Within the framework of the theoretical approach, developed primarily for ``parent'' orthoferrites in the works\,\cite{thesis,JMMM-2016,CM-2019,JETP-2021}, we present the Hamiltonian of the weak ferrimagnet YFe$_{1-x}$ Cr$_x$O$_3$ in the form
\begin{equation}\label{Ham}
\hat H = {\hat H}_{ex} + {\hat H}_{DM} + {\hat H}_{SIA}^{(2)} + {\hat H}_{SIA}^{(4)} + {\hat H}_{TIA}
\end{equation}
of the sum of contributions from isotropic exchange interaction, antisymmetric Dzyaloshinsky-Moriya exchange, single-ion and two-ion spin anisotropy. A rough estimate of the various contributions gives $H_{ex}\sim 1$\,meV, $H_{DM}\sim 0.1$\,meV, $H_{SIA}^{(2)}\sim H_{TIA}\sim 0.01$\,meV, $H_{SIA}^{(4)}\sim 0.001$\,meV.

\subsection{Isotropic superexchange}

Below we restrict ourselves to the bilinear isotropic Heisenberg superexchange nearest neighbor interaction in the form
\begin{equation}\label{Hex}
{\hat H}_{ex} = \frac{1}{2} \sum_{\langle mn \rangle} I_{mn} \, ({\hat {\bf S}}_m \cdot {\hat { \bf S}}_n) ,
\end{equation}
determining the formation of the main $G$-type antiferromagnetic structure, experimentally established using magnetic neutron diffraction for both parent ferrites and weak ferrimagnets of the YFe$_{1-x}$Cr$_x$O$_3$ type (see, for example, \,\cite{Kuznetsov,Dasari,Nair}). The value of the exchange integrals $I_{FeFe}$ and $I_{CrCr}$ can be found from data on the Neel temperatures of YFeO$_3$ and YCrO$_3$ based on a simple mean field approximation
\begin{equation}
	T_N=\frac{zS(S+1)}{3k_B}I \, ,
\end{equation}
although these ``experimental'' exchange integrals ($I_{FeFe}$\,=\,36.8\,K in YFeO$_3$, $I_{CrCr}$\,=\,18.7\,K in YCrO$_3 $), can be one and a half to two times less than those obtained by other methods\,\cite{thesis,LuCrO3,YFeO3-2014}. Thus, recent experimental data give for YFeO$_3$ $I^{(21)}=58.2$\,K, $I^{(41)}=53.6$\,K\,\cite{Park} or $I ^{(21)}\approx I^{(41)}=51.5$\,K\,\cite{Amelin}.

The only source of experimental information on the value of the exchange integral $I_{FeCr}$ in orthoferrites-orthochromites are experimental data\,\cite{Ovanesyan} from a study of the Mössbauer spectra of ${}^{57}$Fe in
RCr$_{0.99}$Fe$_{0.01}$O$_3$ (R $=$ La, La$_{0.5}$Nd$_{0.5}$, Nd, Sm, Gd, Dy, Y, Er, Yb, Lu), according to which in all systems $I_{FeCr}$ has an ``antiferromagnetic sign'' $I_{FeCr} > 0$, decreasing during the transition from ``heavy'' to ``light'' R-ions. Thus, $I_{FeCr} = 13.6 \pm 0.8,\ 13.4 \pm 0.4,\ 7.2 \pm 0.4$ K for R\,=\,Lu, Y, La, respectively.
The model theory of the angular dependence of superexchange integrals well describes the experimental data for the exchange integrals $I_{FeFe}$, $I_{CrCr}$ in orthoferrites and orthochromites\,\cite{JMMM-2016,CM-2019,JETP-2021} and predicts the change sign for $I_{FeCr}$ at the Fe--O--Cr superexchange bond angle $\theta_{12}\approx 170^{\circ}$. In other words, superexchange (Fe$^{3+}$\,--\,O$^{2-}$\,--\,Cr$^{3+}$) becomes ferromagnetic for bond angles significantly exceeding angles typical for orthoferrites-orthochromites ($140^{\circ}-155^{\circ}$).
Unfortunately, at present there are no reliable estimates of ``non-Heisenberg'' isotropic interactions, in particular, biquadratic exchange
\begin{equation}
{\hat H}_{bq}=\frac{1}{2} \sum_{\langle mn \rangle} j_{mn} \, ({\hat {\bf S}}_m \cdot {\hat { \bf S}}_n)^2 \, .
\end{equation}

\subsection{Antisymmetric Dzyaloshinsky-Moriya exchange}

Antisymmetric Dzyaloshinskii-Moriya exchange\,\cite{Dzyaloshinskii,Moriya}
\begin{equation}\label{HDM}
{\hat H}_{DM} = \frac{1}{2} \sum_{\langle mn \rangle} {\bf d}_{mn} \cdot [ {\hat {\bf S}}_m \times {\hat {\bf S}}_n ]\,
\end{equation}
determines, as a rule, the main, antisymmetric contribution to the Dzyaloshinskii interaction\,\cite{Dzyaloshinskii}.
Microscopic theory of bilinear antisymmetric exchange ${\hat H}_{DM}$
was discussed in detail in a number of works by one of the authors\,\cite{1977,thesis,JMMM-2016,CM-2019,JETP-2021}. In particular, he established a microscopic formula for the Dzyaloshinsky vector in a pair of superexchange-coupled $S$ type magnetic ions with an orbitally nondegenerate ground state under the assumption of local cubic symmetry, that is, for $3d$ ions (Fe$^{3+} $, Mn$^{2+}$, Cr$^{3+}$, Mn$^{4+}$, Ni$^{2+}$) with half-filled $t_{2g}$- and/ or $e_g$-shells ($t_{2g}^3$, $t_{2g}^3e_g^2$, $t_{2g}^6e_g^2$)\,\cite{1971,thesis,JMMM-2016,CM-2019,JETP-2021}
\begin{equation}
	\mathbf{d}_{mn} = d_{mn}(\theta) [\mathbf{r}_m \times \mathbf{r}_n] \, ,
	\label{d12}
\end{equation}
where ${\bf r}_{m,n}$ are unit radius vectors for bonds O\,--\,M$_{m,n}$ with presumably equal bond lengths, and the factor $d_{mn}( \theta)$ has a simple angular dependence
\begin{equation}
	d_{mn}(\theta)=d_1+d_2\,\cos\theta_{mn}\, ,
\end{equation}
where $\theta_{mn}$ is the angle connections M$_{m}$\,--\,O\,--\,M$_{n}$. The sign of the scalar parameter $d_{mn}(\theta)$ can be considered as the sign of the Dzyaloshinsky vector. Theoretically predicted sign relationships
Dzyaloshinsky vectors in pairs of $3d$ ions of $S$ type with local octahedral symmetry and superexchange coupling angle $\theta >\theta_{cr}$ ($\cos \theta_{cr}$\,=\,$-d_1 /d_2$) are given in Table\,\ref{tablesign}.

\begin{table}[h]
\begin{center}
\caption{Theoretical predictions of the signs of Dzyaloshinsky vectors in pairs of $3d$ ions of $S$ type
with local octahedral symmetry and superexchange coupling angle $\theta >\theta_{cr}$.}
\begin{tabular}{c|c|c|c}
3$d^n$ & Cr$^{3+}$, Mn$^{4+}$ & Mn$^{2+}$, Fe$^{3+}$ & Ni$^{2+} $, Cu$^{3+}$ \\ \hline
Cr$^{3+}$, Mn$^{4+}$ & + &--& + \\ \hline
Mn$^{2+}$, Fe$^{3+}$ & -- &+ & + \\ \hline	
Ni$^{2+}$, Cu$^{3+}$ & +& +& -- \\
\end{tabular}
\label{tablesign}
\end{center}
\end{table}

The anisotropic antisymmetric Dzyaloshinsky-Moriya interaction, the magnitude and orientation of the Dzyaloshinsky vectors, is responsible for the formation of relatively weak overt (F-mode) and hidden (A and C-mode) spin noncollinearity in the model of four magnetic sublattices, that is, weak ferromagnetism and weak antiferromagnetism, respectively -- ``fine structure'' of three possible magnetic configurations $\Gamma_1\,(A_x, G_y, C_z)$, $\Gamma_2\,(F_x, C_y, G_z)$, $\Gamma_4\,(G_x, A_y, F_z)$. The weak ferromagnetic moment, or the F-mode of noncollinearity, is determined by the $y$-component of the Dzyaloshinsky vectors, the weak antiferromagnetic A- and C-modes are determined by the $z$- and $x$-components of the Dzyaloshinsky vectors, respectively\,\cite{JMMM-2016,CM-2019,JETP-2021}.


In 1975, based on a simple formula for the Dzyaloshinsky vector (\ref{d12}), a connection was established between the crystallographic and ``canted'' magnetic structures for four-sublattice orthoferrites RFeO$_3$ and orthochromites RCrO$_3$\,\cite{FTT-1975,thesis}:
$$
F_x=\frac{(x_1+2z_2)ac}{6l^2}\frac{d}{I}G_z;\ F_z=-\frac{(x_1+2z_2)ac}{6l^2}\frac{d }{I}G_x\,;\
$$
$$
A_x=\frac{(\frac{1}{2}+y_2-x_2)ab}{2l^2}\frac{d}{I}G_y;\ A_y=-\frac{(\frac{1}{ 2}+y_2-x_2)ab}{2l^2}\frac{d}{I}G_x;
$$
\begin{equation}
C_y=\frac{(\frac{1}{2}-y_1)bc}{2l^2}\frac{d}{I}G_z;\ C_z=-\frac{(\frac{1}{2} -y_1)bc}{2l^2}\frac{d}{I}G_y\,;
\label{FCA}	
\end{equation}
where $a,b,c$ are unit cell parameters, $x_{1,2}, y_{1,2}, z_2$ are oxygen (O$_{I,II}$) parameters, $l$ -- average cation-anion bond length. These relations assume averaging over Fe$^{3+}$\,--\,O$^{2-}$\,--\,Fe$^{3+}$ bonds in the $ab$ plane and along $c$ axis. It is obvious that in this approximation, the relative magnitude of the components of the ``small'' basis vectors that describe the explicit and hidden non-collinearity of the spins is determined only by the structural parameters.

Taking into account (\ref{FCA}), we arrive at a simple relationship between crystallographic parameters and the magnetic moment of the Fe sublattice: in units of emu/g
\begin{equation}
	M_{Fe}=\frac{4g\mu_BS}{\rho V}|F_{x,z}|=\frac{2g\mu_BSac}{3l^2\rho V}(x_1+2z_2)\frac{d (\theta)}{I(\theta)}\, ,
\end{equation}
where $\rho$ and $V$ are the density and volume of the unit cell, respectively.
The explicit skew of magnetic sublattices, or the ferromagnetism vector $F_{x,z}$, can be calculated through the ratio of the Dzyaloshinsky field ($H_D$) to the exchange field ($H_E$)
\begin{equation}
	F=H_D/2H_E \, .
\end{equation}
 If we know the Dzyaloshinsky field, we can calculate the factor $d(\theta )$ in orthoferrites as follows
\begin{equation}
	H_D=\frac{S}{g\mu_B}\sum_i|d_y(1i)|=\frac{S}{g\mu_B}(x_1+2z_2)\frac{ac}{l^2}|d(\theta )|\, ,
\end{equation}
which gives $|d(\theta )|\cong3.2$\,K\,$\approx 0.3$\,meV in YFeO$_3$ at $H_D=140$\,kOe\,\cite{Jacobs}. This value is in good agreement with data from recent experiments\,\cite{Park,Amelin}, which made it possible to find information about the Dzyaloshinsky vector based on measurements of the spin wave spectrum.
Note that despite the fact that $F_z\approx 0.01$ the parameter $d(\theta)$ is only an order of magnitude smaller than the exchange integral in YFeO$_3$.

The connection between the orientation of the Dzyaloshinsky vector and the geometry of superexchange of magnetic cations (\ref{d12}) made it possible to find the relationship between the quantities $F_{x,z}, A_{x,y}, C_{y,z}$, which determine the weak non-collinearity of spins in configurations $\Gamma_1$, $\Gamma_2$, $\Gamma_4$, which is in good agreement with experimental data from nuclear magnetic resonance and magnetic neutron diffraction\,\cite{JMMM-2016,CM-2019}. Note that $|A_{x,y}|>|F_{x,z}|>|C_{y,z}|$.

\begin{figure}[h]
\centering
\includegraphics[width=0.48\textwidth]{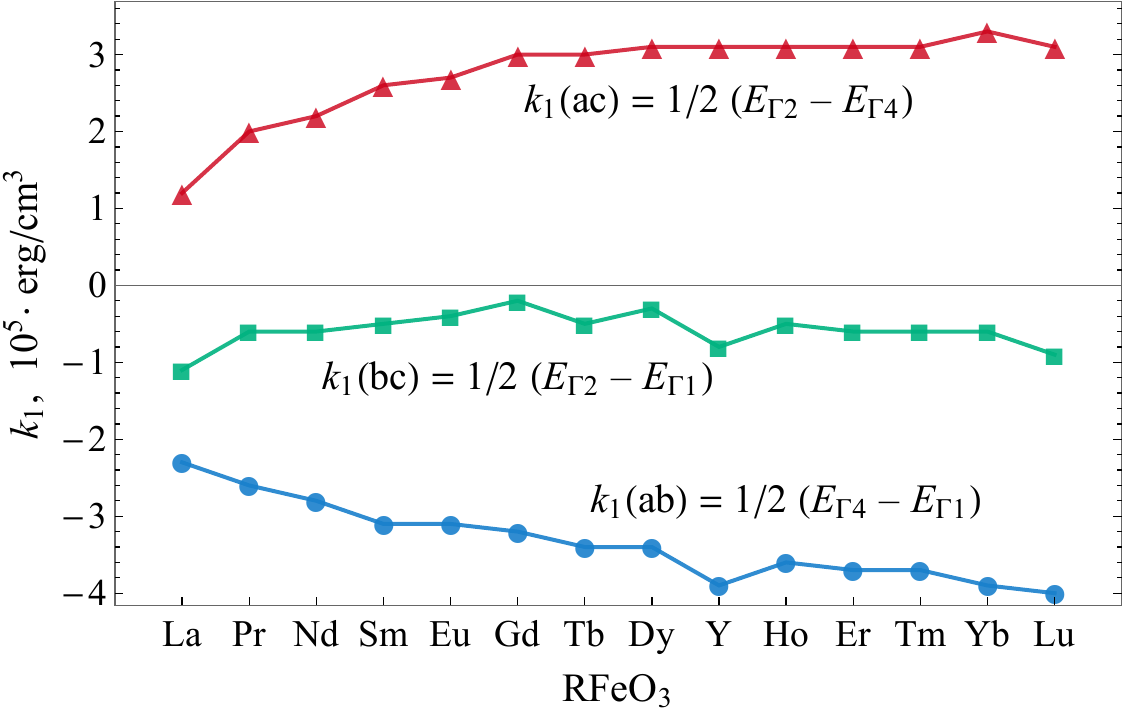}
\caption{
Contribution of antisymmetric exchange to anisotropy constants for different planes in RFeO$_3$
\label{k1(R)}}
\end{figure}

These same relations make it possible to find the contribution of antisymmetric exchange to magnetic anisotropy.
Fig.\,\ref{k1(R)} shows the calculated contribution of antisymmetric exchange to the second-order anisotropy constants for various planes in RFeO$_3$\,\cite{thesis,JMMM-2016,CM-2019,JETP-2021}. These constants determine the effective anisotropy energy $E_{an}=k_1\,\cos2\theta$ for rotation of the antiferromagnetism vector ${\bf G}$ in the $ac$ or $bc$ plane, or $E_{an}= k_1\,\cos2\phi$ for rotating the vector ${\bf G}$ in the $ab$ plane, where $\theta ,\phi$ are the polar and azimuthal angles of the vector ${\bf G}$. It is obvious that antisymmetric exchange stabilizes the $G_x$, or $\Gamma_4$ configuration in all orthoferrites\,\cite{JMMM-2016,CM-2019,JETP-2021} and, most likely, in orthochromites, given the generally similar geometry bonds Cr--O--Cr and Fe--O--Fe, and the energies of the $G_z (\Gamma_2 )$ and $G_y (\Gamma_1 )$ configurations are close. In other words, the contribution to magnetic anisotropy in the $bc$ plane is small, since the constant $k_1(bc) = k_1(ab)+k_1(ac)$ is almost an order of magnitude smaller than the constants $k_1(ac)$ and $k_1(ab )$.

\subsection{ Symmetrical spin anisotropy }

Single-ion spin anisotropy $H_{SIA}$ includes terms of the second and fourth order: $H_{SIA}=H_{SIA}^{(2)}+H_{SIA}^{(4)}$. Due to the low symmetry of the local environment of $3d$ ions in the structure of orthoferrites and orthochromites, we represent second-order single-ion anisotropy in the most general form as
\begin{equation}\label{HSIA2}
{\hat H}_{SIA}^{(2)} = \sum_{n} K_n \Big[ ({\hat {\bf S}}_n\cdot{\bf a}_n) ({\hat { \bf S}}_n\cdot{\bf b}_n)-\frac{1}{3}({\bf a}_n\cdot {\bf b}_n){\hat {\bf S}}_n ^2 \Big]\, ,
\end{equation}
where $K_n$ are some anisotropy parameters, ${\bf a}_n, {\bf b}_n$ are unit vectors that define two characteristic preferred directions for $n$ site, the second term ensures the ``traceless'' of the tensor anisotropy constants. For uniaxial anisotropy ${\bf a}_n = {\bf b}_n$.

Often, especially in spin resonance theory, the expression $H_{SIA}^{(2)}$ is used for an individual ion in the form
\begin{gather}
{\hat H}_{SIA}^{(2)} = D({\hat S}_z^2-\frac{1}{3}S(S+1))+E({\hat S} _x^2-{\hat S}_y^2) + \nonumber\\
+\frac{p}{2}\{{\hat S}_y,{\hat S}_z\}+\frac{q}{2}\{{\hat S}_x,{\hat S}_z \}+\frac{r}{2}\{{\hat S}_x,{\hat S}_y\}\, ,
\label{HSIA2a}
\end{gather}
$D,\ E,\ p,\ q,\ r$ are anisotropy constants, $\{{\hat S}_x,{\hat S}_y\},\ \dots$ are anticommutators.
The first two terms in ${\hat H}_{SIA}^{(2)}$, or the so-called ``rhombic'' anisotropy, determine the contribution to the first anisotropy constants for $ac$, $ab$, and $bc$ planes in orthoferrites and orthochromites. The last three terms determine the symmetric contribution to the Dzyaloshinsky interaction, taking into account which we obtain a symmetric contribution to the explicit and hidden noncollinearity of spins:
\begin{equation}
F_{x,z}=\frac{q}{6I}G_{z,x}\,;\ A_{x,y}=\frac{r}{4I}G_{y,x}\,;\ C_{y,z}=\frac{p}{2I}G_{z,y} \,.
\label{FCAsym}	
\end{equation}

Naturally, taking into account the signs in the relationships between small basis vectors ${\bf F},{\bf A},{\bf C}$ with the ``large'' antiferromagnetism vector ${\bf G}$ in (\ref {FCA}) and (\ref{FCAsym}) antisymmetric and symmetric contributions add or subtract during reorientation of the $G_x\leftrightarrow G_z$ type, in particular, thereby violating the orthogonality of the ferro- and antiferromagnetism vectors (${\bf F}$ and ${\bf G}$) in the angular $G_{xz}$ phase, and leading to a change in the magnetization value during spin reorientation.
Both theoretical estimates\,\cite{thesis} and experimental ESR data of Fe$^{3+}$ and Cr$^{3+}$ ions in LaAlO$_3$ orthoaluminates isostructural with orthoferrites and orthochromites\,\cite{Taylor} and YAlO$_3$\,\cite{White} indicate that the values of the parameters $p,q,r$ do not exceed 0.1\,K.

Taking into account the experimental values of exchange integrals, we expect for a ``symmetric'' contribution to the ferromagnetism vector for orthoferrite YFeO$_3$ and orthochromite YCrO$_3$: $|F_{x,z}|\leq 5\cdot 10^{-4 }$ and $|F_{x,z}|\leq 10^{-3}$ respectively, that is, 5 and 10\% of the value of $F\approx 0.01$, typical for the main ``antisymmetric'' contribution. Let's note, that experimental estimates of work\,\cite{Jacobs} give slightly lower estimates, 2 and 5\%, respectively. It is obvious that for weak ferrimagnets of the YFe$_{1-x}$Cr$_x$O$_3$ type, especially in the region of compensation of the main ``antisymmetric'' contribution to magnetization, the small ``symmetric'' contribution of single-ion anisotropy can play an important role, making a noticeable contribution to the total magnetic moment. At the same time, the possibility of competition between the signs of the contributions of the Fe and Cr sublattices cannot be excluded.

Second-order single-ion spin anisotropy turns out to be very sensitive to distortions in the local environment of $3d$ ions. Thus, in the orthoferrites YFeO$_3$ and LuFeO$_3$, the contribution $H_{SIA}^{(2)}$ to the first anisotropy constant in the $ac$ plane $k_1(ac)$ differs in sign, so that in YFeO$_3$ this contribution partially compensates for the contribution of the Dzyaloshinsky-Moriya interaction, and in LuFeO$_3$ both contributions add up, resulting in a sharp increase in $k_1(ac)$\,\cite{JETP-1981}. The difference in the electronic structure of the Fe$^{3+}$($t_{2g}^3e_g^2$) and Cr$^{3+}$($t_{2g}^3$) ions leads to a significantly different reaction of the contribution of the single-ion second-order anisotropy on local structure distortions. It is possible that the sharp change in the $H_{SIA}^{(2)}$ contribution is responsible for the transition from the basic $\Gamma_4$ configuration in YFeO$_3$ to the $\Gamma_2$ configuration in LuCrO$_3$\,\cite{LuCrO3}.
Let us pay attention to the close value, but different sign of the effective second-order single-ion anisotropy constant for the impurity ions Fe$^{3+}$ and Cr$^{3+}$ in the isostructural to orthoferrites and orthochromites orthoaluminate LaAlO$_3$\,\cite{Taylor}.

In the fourth-order single-ion anisotropy, which is nonzero only for Fe$^{3+}$ ions ($S=5/2$), we highlight the main term, which in the local coordinate system for weak distorted FeO$_6$ octahedra in orthoferrites can be represent in the form
\begin{equation}\label{HSIA4}
{\hat H}_{cub} = \frac{a}{6} \sum_n \Big[ {\hat S}_{x_n}^4 + {\hat S}_{y_n}^4 + {\hat S}_{z_n}^4 - \frac{1}{5} S (S + 1)(3S^2 + 3S - 1) \Big] \, ,
\end{equation}
the so-called ``cubic'' anisotropy\,\cite{cubic,M-2021}. By the way, the value $a = 4\cdot 10^{-3}$\,cm$^{-1}$, obtained from ESR data of Fe$^{3+}$ ions in LaAlO$_3$\,\cite{Taylor} can be considered a reasonable estimate of the cubic anisotropy constant of Fe$^{3+}$ ions and in YFe$_{1-x}$Cr$_x$O$_3$.

Two-ion anisotropy $H_{TIA}$ includes both the classical bilinear magnetic-dipole interaction and more complex exchange-relativistic contributions\,\cite{TIA}. The relatively small contribution of the magnetic dipole interaction, as well as the contribution of $H_{DM}$, stabilizes the magnetic configuration of $\Gamma_4$, and its value, decreasing with the transition to ``light'' R-ions, practically becomes negligible in LaFeO$_3$\,\cite{thesis,M-2021}.

\section{Molecular field models for weak ferrimagnets of the YF\lowercase{e}$_{1-x}$C\lowercase{r}$_x$O$_3$ type}

The molecular field approximation (MFA) provides the basis for the simplest and most physically visual, albeit semi-quantitative, description of spin systems. In many cases, especially for such complex systems as weak ferrimagnets such as YFe$_{1-x}$Cr$_x$O$_3$, the advantage of this method is the ability to describe the fundamental features of the concentration and temperature dependences of the main magnetic characteristics, i.e. the Neel temperature, magnetization, magnetic anisotropy.

However, it should be taken into account that the approximations used in any MFA method can significantly affect the results. To demonstrate this, below we will consider two scheme of the molecular field theory for describing bilinear interactions, i.e. isotropic exchange and antisymmetric Dzyaloshinsky-Moriya interaction in weak ferrimagnets such as YFe$_{1-x}$Cr$_x$O$_3$.

We also note that the classical Monte Carlo (MC) method\,\cite{Billoni} has an advantage over MFA in the ability to consistently take into account all possible configurations of the distribution of Fe and Cr ions in the lattice, but unlike MFA within the framework of the classical spin MC method ions are replaced by classical vectors, which undoubtedly introduces significant errors into the results.

\subsection{MFA-I}

In the simplest model of weak ferrimagnets of the RFe$_{1-x}$Cr$_x$O$_3$ type (R = La, Y, Lu), which assumes a single magnetic ordering in the Fe--Cr subsystem with molecular fields common to all Fe$^{3+}$ (Cr$^{3+}$) ions, the bilinear part of the Hamilton operator is represented as
\begin{equation}\label{HMFA}
{\hat H}_{ex} + {\hat H}_{DM} = \sum_n\,({\bf h}_n\cdot{\hat {\bf S}}_n) -\frac{1} {2}\sum_n\,({\bf h}_n\cdot \langle{\bf S}_n\rangle ) \, ,
\end{equation}
where for the molecular field ${\bf h}_n$, taking into account the leading contributions of isotropic exchange and the Dzyaloshinsky-Moriya interaction, we have
\begin{equation}\label{MFAh}
{\bf h}_n = \sum_m\,\left(I_{mn} \langle{\bf S}_m\rangle + [{\bf d}_{mn}\times \langle{\bf S}_m\rangle ]\right)\, ,
\end{equation}
where $\langle{\bf S}_m\rangle $ is the thermodynamic average spin of an arbitrary ion (Fe$^{3+}$ or Cr$^{3+}$)
\begin{equation}
\langle {\bf S}_m\rangle = -\frac{{\bf h}_m}{h_m}\,S\,B_S \left( \frac{Sh_m}{k_B T} \right) \, ,
\label{MFAeq}
\end{equation}
where $B_S$ is the function Brillouin, ${\bf h}_m$ is an effective field for $m$-th site, $h_m=|{\bf h}_m|$.

It is obvious that, in contrast to the homogeneous parent systems YFeO$_3$ and YCrO$_3$, for weak ferrimagnets of the YFe$_{1-x}$Cr$_x$O$_3$ type we are forced to introduce a number of additional assumptions and approximations to solve the molecular field equations (\ref{MFAeq}):

1) Fe$^{3+}$ and Cr$^{3+}$ ions occupy lattice sites with equal probability;

2) the parameters of the spin Hamiltonian do not depend on either the local configuration or the concentration of Fe$^{3+}$ and Cr$^{3+}$ ions;

3) long-range crystalline and magnetic (spin) order is preserved, in other words, the classification of possible magnetic structures ($\Gamma_{1,2,4}$) and the corresponding relationships between the average values of spin moments in positions 1, 2, 3 and 4 are also preserved (see Table\,\ref{GammaTable}), which allows us to consider the molecular field equations only for one position of $3d$ ions.

\begin{table}[h]
\begin{center}
\caption{Connection between spin components on different sublattices in phases $\Gamma_1 (A_x, G_y, C_z)$, $\Gamma_2 (F_x, C_y, G_z)$, $\Gamma_4 (G_x, A_y, F_z)$}\label{GammaTable}
\def\arraystretch{1.5}
\begin{tabular}{c | c}
& $ S_{x}^{(1)} = -S_{x}^{(2)} = -S_{x}^{(3)} = S_{x}^{(4)} $ \\
$\Gamma_1$ & $ S_{y}^{(1)} = -S_{y}^{(2)} = S_{y}^{(3)} = -S_{y}^{(4) } $\\
& $ S_{z}^{(1)} = S_{z}^{(2)} = -S_{z}^{(3)} = -S_{z}^{(4)} $ \\
\hline
& $ S_{x}^{(1)} = S_{x}^{(2)} = S_{x}^{(3)} = S_{x}^{(4)} $ \\
$\Gamma_2$ & $ S_{y}^{(1)} = S_{y}^{(2)} = - S_{y}^{(3)} = - S_{y}^{(4) } $\\
& $ S_{z}^{(1)} = - S_{z}^{(2)} = S_{z}^{(3)} = - S_{z}^{(4)} $ \\
\hline
& $ S_{x}^{(1)} = - S_{x}^{(2)} = S_{x}^{(3)} = - S_{x}^{(4)} $ \\
$\Gamma_4$ & $ S_{y}^{(1)} = - S_{y}^{(2)} = - S_{y}^{(3)} = S_{y}^{(4) } $ \\ & $ S_{z}^{(1)} = S_{z}^{(2)} = S_{z}^{(3)} = S_{z}^{(4)} $
\end{tabular}
\end{center}
\end{table}

Thus, for the molecular field ${\bf h}_{Fe}$ in position 1 we obtain
\begin{multline}
{\bf h}_{Fe} = P_{Fe}(x) \Big\langle 4 I_{FeFe}\, {\bf \hat{S}}^{(4)}_{Fe} + 2 I_ {FeFe}\, {\bf \hat{S}}^{(2)}_{Fe} + \\
+ 4 [ {\bf d}^{(41)}_{FeFe} \times {\bf \hat{S}}^{(4)}_{FeFe} ] + 2 [ {\bf d}^{ (21)}_{FeFe} \times {\bf \hat{S}}^{(2)}_{FeFe} ] \Big\rangle + \\
+ P_{Cr}(x) \Big\langle 4 I_{FeCr}\, {\bf \hat{S}}^{(4)}_{Cr} + 2 I_{FeCr}\, {\bf \ hat{S}}^{(2)}_{Cr} + \\
+ 4 [ {\bf d}^{(41)}_{FeCr} \times {\bf \hat{S}}^{(4)}_{Cr} ] + 2 [ {\bf d}^{ (21)}_{FeCr} \times {\bf \hat{S}}^{(2)}_{Cr} ] \Big\rangle ,
\label{hFe14}
\end{multline}
where $P_{Fe}(x) = 1 - x$, $P_{Cr}(x) = x$ are the concentrations of Fe$^{3+}$ and Cr$^{3+}$ ions, respectively, the brackets $\langle \dots \rangle$ denote the thermodynamic mean, and the components of the vectors ${\bf S}^{(2)}$ and ${\bf S}^{(4)}$ are expressed in terms of ${\bf S }^{(1)}$ in accordance with the Table \ref{GammaTable}. The field ${\bf h}_{Cr}$ is obtained by replacing FeFe and FeCr with, respectively, FeCr and CrCr on the right side. Let us note that in (\ref{hFe14}) nonequivalent contributions of bonds 1-2 and 1-4 are highlighted, which is especially important given the different orientation and magnitude of the Dzyaloshinsky vectors for these bonds.

The works\,\cite{Dasari,Billoni} consider the classical approximation for weak ferrimagnets of the YFe$_{1-x}$Cr$_x$O$_3$ type with ``limiting'' averaging over configurations, in which, for example, the Hamiltonian of isotropic exchange is represented as energy (per ion in position 1)
\begin{multline}
E = z \sum_{\alpha\beta} P_{\alpha}(x)P_{\beta}(x) I_{\alpha\beta} \, ({\bf S}_{\alpha} \cdot { \bf S}_{\beta}) = \\
= z \big[ (1-x)^2 I_{FeFe}({\bf S}_{Fe} \cdot {\bf S}_{Fe}) + x^2 I_{CrCr} ({\bf S}_{Cr} \cdot {\bf S}_{Cr}) + \\
+2x(1-x) I_{FeCr} ({\bf S}_{Fe} \cdot {\bf S}_{Cr}) \big] \, .
\label{Dasari}
\end{multline}

\subsection{MFA-II}

A more accurate version of the molecular field model for weak ferrimagnets (MFA-II), which takes into account the dependence of small components of the local molecular field, which describe the effects of overt (F-type) and hidden (A-, C-type) spin non-collinearity, on the configuration of the nearest environment, was proposed and implemented earlier in the works\,\cite{1977,thesis,YLuFeCrO3,APP-1985,FeCrBO3,MnNiCO3}.

Let us consider the Fe$^{3+}$ ion in the $S^{(1)}$ position; for the Cr$^{3+}$ ion the formulas will be similar. Depending on the configuration of the environment, the molecular fields will have the following form:
\begin{multline}\label{meanfII}
{\bf h}_{Fe}^{klr} = k\, \left(I_{FeFe}\, \langle {\bf S}^{(4)}_{Fe} \rangle + \big[{ \bf d}^{(41)}_{FeFe} \times \langle {\bf S}^{(4)}_{Fe} \rangle\big] \right) + \\
+l\, \left(I_{FeFe}\, \langle {\bf S}^{(4)}_{Fe} \rangle + \big[{\bf d}'^{(41)}_{ FeFe} \times \langle {\bf S}^{(4)}_{Fe} \rangle\big] \right) + \\
+ r\, \left(I_{FeFe}\, \langle {\bf S}^{(2)}_{Fe} \rangle + \big[{\bf d}^{(21)}_{FeFe } \times \langle {\bf S}^{(2)}_{Fe} \rangle\big] \right) + \\
+ (u - k)\, \left(I_{FeCr}\, \langle {\bf S}^{(4)}_{Cr} \rangle + \big[{\bf d}^{(41) }_{FeCr} \times \langle {\bf S}^{(4)}_{Cr} \rangle\big] \right) + \\
+ (u - l)\, \left(I_{FeCr}\, \langle {\bf S}^{(4)}_{Cr} \rangle + \big[{\bf d}'^{(41 )}_{FeCr} \times \langle {\bf S}^{(4)}_{Cr} \rangle\big] \right) + \\
+ (u - r)\, \left(I_{FeCr}\, \langle {\bf S}^{(2)}_{Cr} \rangle + \big[{\bf d}^{(21) }_{FeCr} \times \langle {\bf S}^{(2)}_{Cr} \rangle\big] \right) ,
\end{multline}
where $k$, $l$ and $r$ are the number of nearest neighbors of type Fe$^{3+}$, and $0 \leq k \leq u$, $0 \leq l \leq u $ and $0 \leq r \leq u$, respectively, where $u = 2$. The need to use three numbers $k$, $l$ and $r$ is due to the fact that the $S^{(1)}$ ion is connected to its neighbors by three different Dzyaloshinsky vectors (see Table \ref{rxr}). The number of neighbors of type Cr$^{3+}$ is determined similarly by the inverse numbers $u - k$, $u - l$ and $u - r$.

\begin{table}[h]
\begin{center}
\caption{ Components $x,\ y,\ z$ of structural factors $[{\bf r}_m \times {\bf r}_n]$ that determine orientation vectors Dzyaloshinsky in YFe$_{0.5}$Cr$_{0.5}$O$_3$, calculated by data neutron diffraction \,\cite{Yang}} \label{rxr}
\begin{tabular}{ c | ccc}
& $x$ & $y$ & $z$ \\
\hline
$ [{\bf r}_2 \times {\bf r}_1]$ & $0.216$ & $0.562$ & $0$ \\
$ [{\bf r}_4 \times {\bf r}_1]$ & $\pm 0.303$ & $0.287$ & $0.397$
\end{tabular}
\end{center}
\end{table}
 
Thus, any average value, such as the magnetic moment, is calculated according to the following rule:
\begin{multline}
\langle {\bf S}_{Fe} \rangle = \sum_{k = 0}^{u}\sum_{l = 0}^{u}\sum_{r = 0}^{u} p^{ klr} \, \langle {\bf S}_{Fe}^{klr} \rangle = \\
=- \sum_{klr} p^{klr} \, \frac{{\bf h}_{Fe}^{klr}}{h_{Fe}^{klr}}\,S_{Fe}\,B_ {S_{Fe}} \left( \frac{S_{Fe} h_{Fe}^{klr}}{k_B T} \right) \, ,
\label{moments2MFAII}
\end{multline}
where $p^{klr} = \binom{u}{k} \binom{u}{l} \binom{u}{r} (1-x)^{k + l + r} x^{z - k - l - r}$ is the probability of occurrence of a configuration with numbers $k$, $l$ and $r$ at concentration $x$, $\binom{u}{l}$ is the binomial coefficient, $z=6 $. Here and below, summation $\sum_{klr}$ will be denoted by $\sum_{k = 0}^{u}\sum_{l = 0}^{u}\sum_{r = 0}^{u}$.

Equations (\ref{moments2MFAII}) can be reduced to linear form if in (\ref{meanfII}) we discard small terms and leave only those that contain exchange $I$ or a component of the vector ${\bf G}$ in the corresponding phase. In most cases, such an approximation does not noticeably affect the results, but, for example, for the vector ${\bf C}$ in phase $\Gamma_2$ the error can reach 10\%.

Taking into account the assumptions made, the equations for the transverse components of magnetic moments in the $\Gamma_4$ phase have the form
\begin{gather}
m^z_{Fe} (\alpha_{Fe}^{+} + 1) + m^z_{Cr} \beta_{Fe}^{+} = -\gamma_{Fe}^F\,; \nonumber\\
m^z_{Fe} \alpha_{Cr}^{+} + m^z_{Cr} (\beta_{Cr}^{+} + 1) = -\gamma_{Cr}^F \,; \nonumber\\
m^y_{Fe} (\alpha_{Fe}^{-} + 1) + m^y_{Cr} \beta_{Fe}^{-} = \gamma_{Fe}^A\,; \nonumber\\
m^y_{Fe} \alpha_{Cr}^{-} + m^y_{Cr} (\beta_{Cr}^{-} + 1) = \gamma_{Cr}^A \,.
\end{gather}
The variable parameters $\alpha$, $\beta$, $\gamma$ are given in Appendix \ref{MFAIIparams}.

\subsection{Effective field method for taking into account single-ion anisotropy}

To calculate the energy of single-ion anisotropy, you can use the effective field approximation\,\cite{M-2021}, in which the second-order spin anisotropy energy for an individual ion takes the form
\begin{gather}
E_{SIA}^{(2)} = D(T) \Big( \cos^2\theta -\frac{1}{3} \Big) + E(T) \sin^2 \theta \cos 2 \phi + \nonumber\\
+ \frac{p(T)}{2} \sin 2\theta \sin \phi + \frac{q(T)}{2} \sin 2\theta \cos\phi +\frac{r(T) }{2} \sin^2\theta \sin 2\phi\, ,
\label{HSIA2b}
\end{gather}
Moreover, the temperature dependence of all anisotropy constants is determined by a single factor $\rho_{2}(T)$:
\begin{gather}
\rho_{2}(T) = \frac{\langle 3\hat{S}_{z}^{2}-S(S+1) \rangle}{S(2S-1)} \, ,
\end{gather}
where the thermodynamic average appears, calculated taking into account the molecular field acting on the ion.

The ``cubic'' contribution to the energy of fourth-order anisotropy can be represented as\,\cite{cubic,M-2021}
\begin{equation}
 	E_{cub} = k_{cub}(T)C^{4A_1}_{0}({\bf S})\,,
 	\label{kA2}
\end{equation}
Where 
$$
C^{4A_1}_{0}({\bf S})=\sqrt{\frac{7}{12}}C^{4}_{0}({\bf S})+\sqrt{\frac{5}{24}} \Big[C^{4}_{4}({\bf S})+C^{4}_{-4}({\bf S}) \Big]
$$
is the so-called invariant cubic tensor harmonic, the argument of which is the orientation angles $\theta ,\phi$ of the classical ion spin vector,
\begin{widetext}
\begin{equation}
k_{cub}(T)= k_{cub}(0)\frac{\langle 35 \hat{S}_z^4-(30S^2+30S-25) \hat{S}_z^2 +3S^ 4+6S^3-3S^2-6S\rangle}{2\sqrt{(2S+5)(2S+3)(2S+1)(2S-1)(2S-3)(S+2)( S+1)S(S-1)}}
\label{kA3}
\end{equation}
\end{widetext}
is the temperature-dependent cubic anisotropy constant, with $a=\,\frac{5\sqrt{3}}{12}k_{cub}(0)$ at $S=5/2$.

In practice, it is important to take into account cubic spin anisotropy for different crystal planes by replacing ${\bf S}\rightarrow {\bf G}$ in the argument of the spherical harmonic, which limits the rotation of the $\bf G$ vector in a certain plane and highlighting the fourth-order contribution:
\begin{equation}
\Phi_{an}^{(4)}=k_2\,\cos4\theta
\end{equation}
for $ac$, $bc$ planes or
\begin{equation}
\Phi_{an}^{(4)}=k_2\,\cos4\varphi
\end{equation}
for the $ab$ plane.

In general, the constants $k_2$ in RFeO$_3$ decrease quite smoothly in absolute value\,\cite{cubic,M-2021}, changing by no more than a factor of two when going from La to Lu. The difference between the constants $k_2(ac)$ and $k_2(bc)$ can serve as a measure of deviation from the ideal cubic perovskite structure, for which $k_2(ac)=k_2(bc)=-\frac{ 3}{4}k_2(ab )$. The different signs of these constants, positive for the $ac$, $bc$ planes and negative for the $ab$ plane, indicate the different nature of spin-reorientation transitions in the corresponding planes, that is, second-order transitions in the $ac$, $ bc$ planes and first-order transitions in the $ab$ plane. Indeed, all currently known spin-reorientation transitions of the $\Gamma_4 - \Gamma_2$ ($G_x - G_z$) type in RFeO$_3$ orthoferrites (R = Sm, Nd, Er, Tm) are smooth, with two characteristic temperatures phase transitions of the second order, which are the beginning and end of spin reorientation, and the only transition of the $\Gamma_4 - \Gamma_1$ ($G_x - G_y$) type (DyFeO$_3$) known for these crystals is a stepwise transition of the first-order\,\cite{KP}. A unique example that confirms our conclusions about the sign of the second anisotropy constant is the mixed orthoferrite Ho$_{0.5}$Dy$_{0.5}$FeO$_3$\,\cite{KP} in which two spin-reorientation transitions $G_x - G_y$ ($T= 46$\,K) and $G_y - G_z$ ($T = 18 \div 24$\,K), realized through one first-order phase transition in the $ab$ plane and two second-order phase transitions in the $bc$ plane, respectively\,\cite{KP}.

\section{Neel temperature of weak ferrimagnets}
\begin{figure}[h]
\centering
\includegraphics[width=0.48\textwidth]{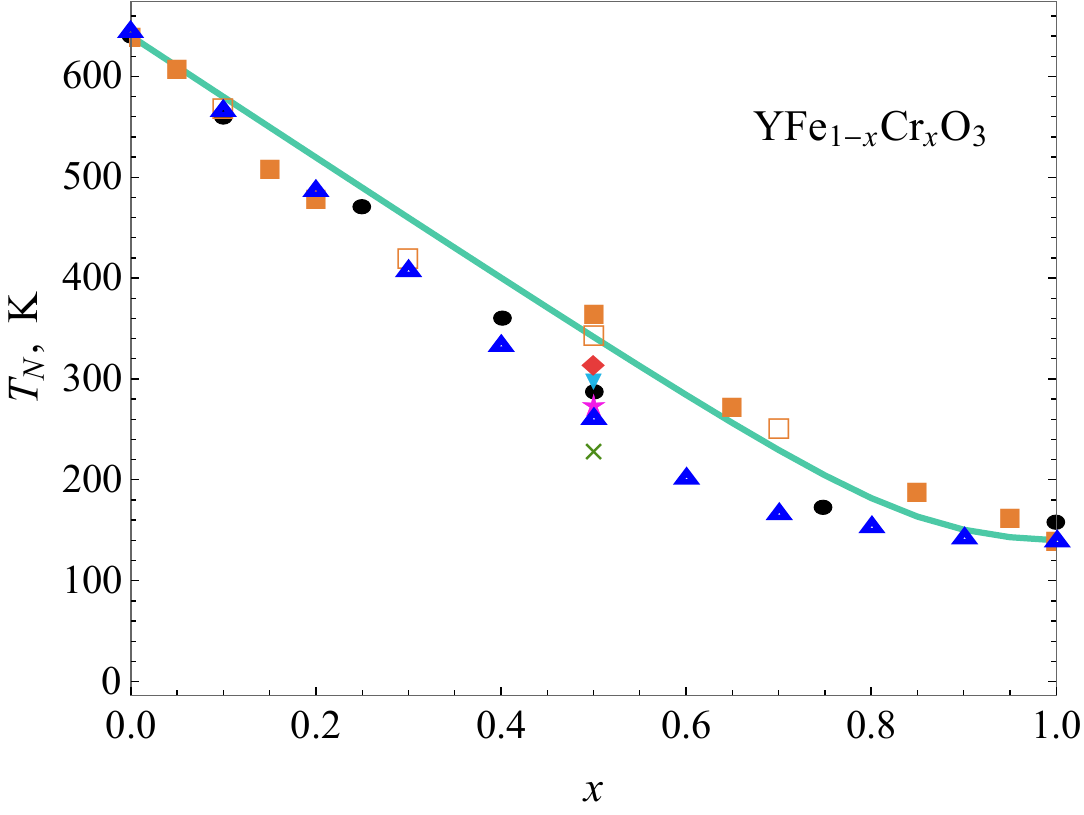}
\caption{
Concentration dependence of the Neel temperature in a weak ferrimagnet YFe$_{1-x}$Cr$_x$O$_3$. Experiment: $\square$ and $\blacksquare$ -- \cite{WFIM-1,WFIM-1976} (poly- and single crystals, respectively); $\blacklozenge$ -- \cite{Yang}, $\blacktriangledown$ -- \cite{Yin2017}, $\bullet$ -- \cite{Salazar2022}, $\star$ -- \cite{Nair}, $\blacktriangle$ -- \cite{Dasari,multi,nomulti}, $\times$ -- \cite{Dahmani2002}. Theoretical curve -- molecular field model (see Hashimoto formula (\ref{TNFeCr}))
\label{TN}}
\end{figure}

MFA methods can be used to calculate the concentration dependence of the Neel temperature of weak ferrimagnets.
Fig.\,\ref{TN} shows such dependences for YFe$_{1-x}$Cr$_x$O$_3$ according to data from various authors. First of all, we note that only the first experimental studies of weak ferrimagnets in the laboratory of A.M. Kadomtseva were performed on single-crystalline samples obtained in the laboratory of B. Wanklyn, while all later studies were performed on polycrystalline samples obtained by various methods\,\cite{Dahmani2002,Dasari,Billoni,Salazar2022}. Let us pay attention to the large scatter of experimental data, especially near the ``half'' composition $x = 0.5$, where the values of $T_N$ from different sources differ by almost 100\,K.
On the one hand, this is due to large fluctuations of local configurations with a large deviation from the parent compositions, on the other hand, to the extremely small value of magnetization for such compositions\,\cite{WFIM-1} and the large contribution of short-range order at $T\geq T_N$.
Some authors argue that an accurate determination of Neel temperatures for these compositions is generally impossible\,\cite{Dahmani2002}.
     
In molecular field theory, the $T_N(x)$ dependence is quite easy to obtain within the framework of a kind of ``zero'' or ``exchange'' approximation, in which we neglect the weak effects of the Dzyaloshinsky-Moriya interaction and symmetric spin anisotropy compared to the large isotropic exchange:
\begin{multline}
{\bf h}_{Fe} = P_{Fe}(x) I_{FeFe} \Big\langle 4\, {\bf \hat{S}}^{(4)}_{Fe} + 2\, {\bf \hat{S}}^{(2)}_{Fe} \Big\rangle + \\
+ P_{Cr}(x) I_{FeCr} \Big\langle 4 \, {\bf \hat{S}}^{(4)}_{Cr} + 2 \, {\bf \hat{S} }^{(2)}_{Cr} \Big\rangle ,
\label{mfa}
\end{multline}
in accordance with (\ref{hFe14}). Using the condition of small exchange molecular fields near $T_N$ and the expansion of the Brillouin function
$$
B_S(x\ll 1)\approx \frac{S+1}{3S}x \, ,
$$
we get instead of (\ref{MFAeq}) system linear equations for average spins $\langle {\bf S}_{Fe}\rangle$ and $\langle {\bf S}_{Cr}\rangle$, the condition solvability which is ratio \,\cite{Hashimoto}
\begin{multline}
\big[T_N(x)-(1-x)T_N(Fe)\big] \big[T_N(x)-xT_N(Cr)\big] = \\
= \frac{z^2}{9}x(1-x)S_{Fe}(S_{Fe}+1)S_{Cr}(S_{Cr}+1)I_{FeCr}I_{CrFe} ,
\label{TNFeCr}
\end{multline}
which can be used either to describe the dependence $T_N(x)$, or to find the value of the integrals of the superexchange interaction Fe--O--Cr (Cr--O--Fe) from the known experimental dependence $T_N(x)$. Naturally, the relation (\ref{TNFeCr}) is satisfied under the assumption that exchange integrals are independent of concentration; in addition, it is natural to assume that $I_{FeCr} = I_{CrFe}$.

Exchange integrals $I_{FeFe} = 36.6$\,K and $I_{CrCr} = 18.7$\,K are found using the well-known MFA expressions for the relationship of exchange integrals with the Neel temperatures of orthoferrite YFeO$_3$ $T_N(0) = 640- 655$\,K\,\cite{Shang} and orthochromite YCrO$_3$ $T_N(1) = 140-159$\,K\,\cite{Salazar2022}.


Using the experimental value of the exchange integral $I_{FeCr} = 13.4 \pm 0.4$\,K, found in the work\,\cite{Ovanesyan}, and assuming $I_{FeCr} = I_{CrFe}$, in Fig.\,\ref{TN} the theoretical dependence $T_N(x)$ is presented, which quite satisfactorily describes the ``older'', but more reliable experimental results obtained on both single-crystalline and polycrystalline samples\,\cite{WFIM-1}. Let us note the systematic deviation of the ``new'' data\,\cite{Dasari} obtained on polycrystalline YFe$_{1-x}$Cr$_x$O$_3$ samples from the experimental data of the work\,\cite{WFIM-1}.

Let us also pay attention to the work\,\cite{Dasari}, in which, to describe their experimental data, the authors used strange expressions for calculating average spin values, where in the argument of the Brillouin functions, instead of natural expressions for effective molecular fields (\ref{mfa}) (see also formula (2) from\,\cite{Dasari}) other expressions have appeared, multiplied by $P_{Fe}(x)$ for ${\bf h}_{Fe}$ and $P_{Cr} (x)$ for ${\bf h}_{Cr}$ respectively. Apparently, the authors motivate this by the need to take into account ``the probabilistic aspect of the occupancy of the site on which the mean field is acting''. In our opinion, this deviation from the classical canons of mean field theory led to a number of incorrect conclusions, for example, the dependence $T_N(x)$, the use of which for processing experimental data led to a nonphysical conclusion about the value of the exchange integral $I_{Fe Cr}\approx 48$\,K (!?), significantly exceeding not only $I_{CrCr}$, but also $I_{FeFe}$.

\section{Orientation and magnitude of the Neel vector in weak ferrimagnets}

The positive (antiferromagnetic) sign and rather large value of the exchange integral $I_{FeCr}$ indicates the preservation of the basic $G$-type antiferromagnetic structure in weak ferrimagnets of the YFe$_{1-x}$Cr$_x$O$_3$ type observed in parent orthoferrites and orthochromites. This is confirmed by both direct neutron diffraction data\,\cite{Dasari,Nair,Pomiro}, and indirect data from magnetostriction measurements\,\cite{WFIM-2}.

\begin{figure}[h]
\centering
\includegraphics[width=0.48\textwidth]{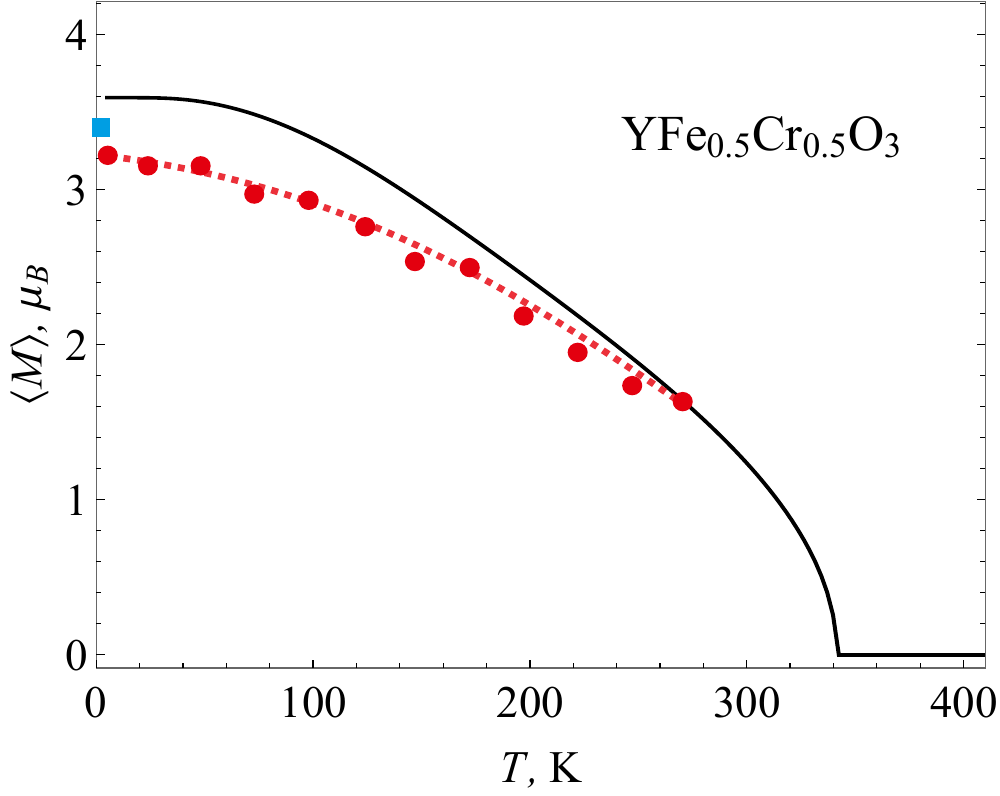}
\caption{
Temperature addiction average magnetic moment in YFe$_{0.5}$Cr$_{0.5}$O$_3$ ($\mu_B/$ion), obtained from neutron diffraction data: $\bullet$ -- \cite{Nair}; $\square$ -- \cite{nomulti}. Solid line -- theory
\label{MFeCr}}
\end{figure}

Fig.\,\ref{MFeCr} shows the temperature dependence of the average magnetic moment
\begin{equation}
\langle M\rangle = g\mu_B \big[ 0.5 \, \langle S_x(Fe)\rangle + 0.5 \, \langle S_x(Cr)\rangle \big]
\end{equation}
in YFe$_{1-x}$Cr$_x$O$_3$ ($x = 0.50$), obtained from neutron diffraction data\,\cite{Nair}, as well as the result of calculation within the MFA-I approximation taking into account $\sim$10\% covalent spin reduction typical of $3d$ oxides. Let us note the completely satisfactory agreement between theory and experiment. Small quantitative discrepancies are naturally explained by the magnetic inhomogeneity of polycrystalline samples.
The data in Fig.\,\ref{MFeCr} indicate a significantly higher real Neel temperature for the samples
YFe$_{0.5}$Cr$_{0.5}$O$_3$ studied in the work\,\cite{Nair} than the value $T_N =275$\,K declared by the authors, based on magnetic measurements under extremely low magnetization, especially since the authors themselves point out the difference between the ZFC and FC curves at temperatures above $275$\,K.

The most reliable results of a neutron diffraction study of the antiferromagnetic $G$-structure of polycrystalline YFe$_{0.5}$Cr$_{0.5}$O$_3$ samples were carried out in a recent work\,\cite{nomulti}, the authors of which discovered an ``angular'' structure with the orientation of the antiferromagnetism vector at $T = 2$\,K at an angle $\approx 22^{\circ}$ to the $a$ axis ($\langle M\rangle =3.4\, \mu_B $, see Fig.\,\ref{MFeCr}), although to determine the exact ``azimuthal'' orientation of the ${\bf G}$ vector in the $bc$ plane, the authors refer to the need for single-crystal samples. These results actually confirmed the ``old'' data from neutron diffraction measurements of the orientation of the antiferromagnetism vector of polycrystalline YFe$_{0.5}$Cr$_{0.5}$O$_3$\,\cite{WFIM-1976} samples, according to which the vector ${ \bf G}$ when the temperature decreases below $260$\,K and down to $6$\,K is oriented at an angle $(30\pm 15)^{\circ}$ to the $a$ axis.

By the way, the authors\,\cite{nomulti} did not find any features of dielectric and pyroelectric properties near the Neel temperature $T_N = 275$\,K declared by them, which, as in the case of the work data\,\cite{Nair}, most likely indicates a significantly higher real Neel temperature in this composition is closer to $T_N = 335$\,K according to the work\,\cite{WFIM-1972}.

Magnetic measurements of the torque curves of single-crystal YFe$_{0.5}$Cr$_{0.5}$O$_3$ samples showed that as the temperature decreases below $T_N = 363$\,K the magnetic moment is unusually small ($\approx 0.022$ emu/g) and up to $T\approx 300$\,K is oriented along the $c$ axis, but with a further decrease in temperature a component of the magnetic moment appears along the $a$ axis, the value of which increases to $\approx 0.17$\, emu/g at $T = 78$\,K. Measurements of torque curves in various planes in a field of 6.25\,kOe showed that at $78 \leq T\leq 171$\,K the total magnetic moment is oriented in the $ac$ plane at an angle of $60^{\circ}$ to $c $ axis, which, generally speaking, does not agree with neutron diffraction data\,\cite{nomulti} for a polycrystal, and indicates an angular phase $G_{xz}$ with the orientation of the antiferromagnetism vector at an angle $60^{\circ}$ to $ a$ axis, but does not exclude a more complex spatial orientation of the antiferromagnetism vector. A similar situation with a ``flat'' angular or spatial orientation of the antiferromagnetism vector was observed in single crystals of almost all compositions YFe$_{1-x}$Cr$_x$O$_3$ ($x =0.15$, 0.38, 0.5, 0.65, 0.85)\,\cite{WFIM-1}, and also, according to the authors of the work\,\cite{WFIM-1976}, and in polycrystalline samples at $x=0.3$, 0.5, 0.7. In all cases, one must also remember the small contribution of single-ion anisotropy to the magnetization; taking this into account at a small total magnetization value disrupts the relationship between the orientation of the magnetic moment and the antiferromagnetism vector, which is typical for the antisymmetric contribution to the Dzyaloshinsky interaction.

\begin{figure}[h]
\centering
\includegraphics[width=0.48\textwidth]{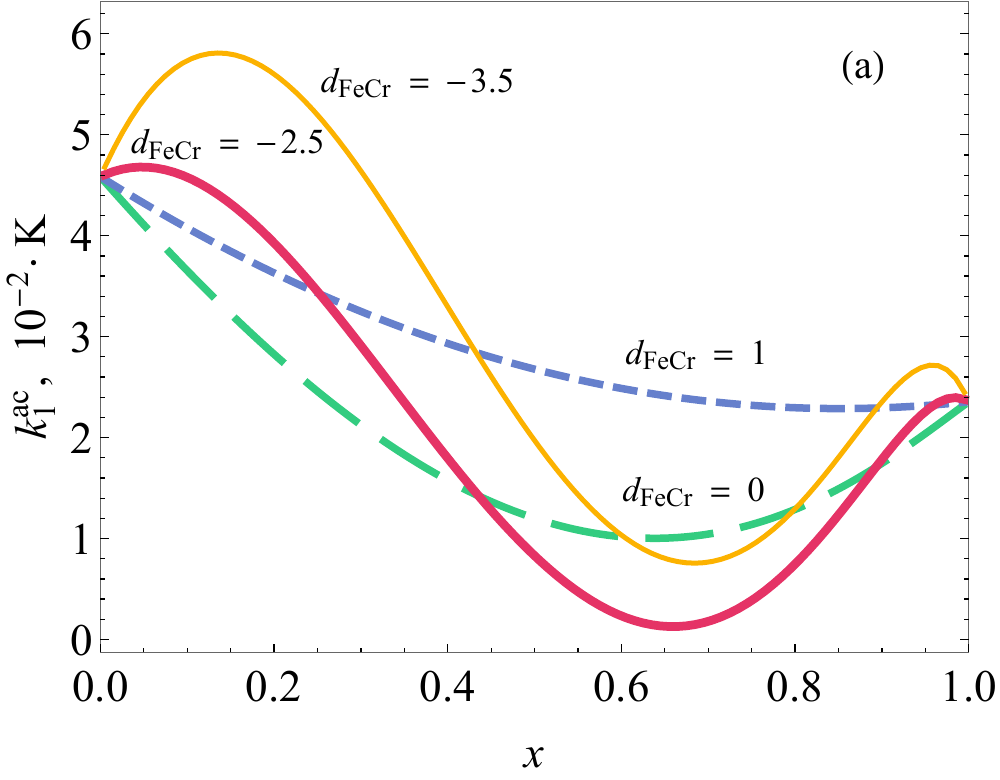}
\includegraphics[width=0.48\textwidth]{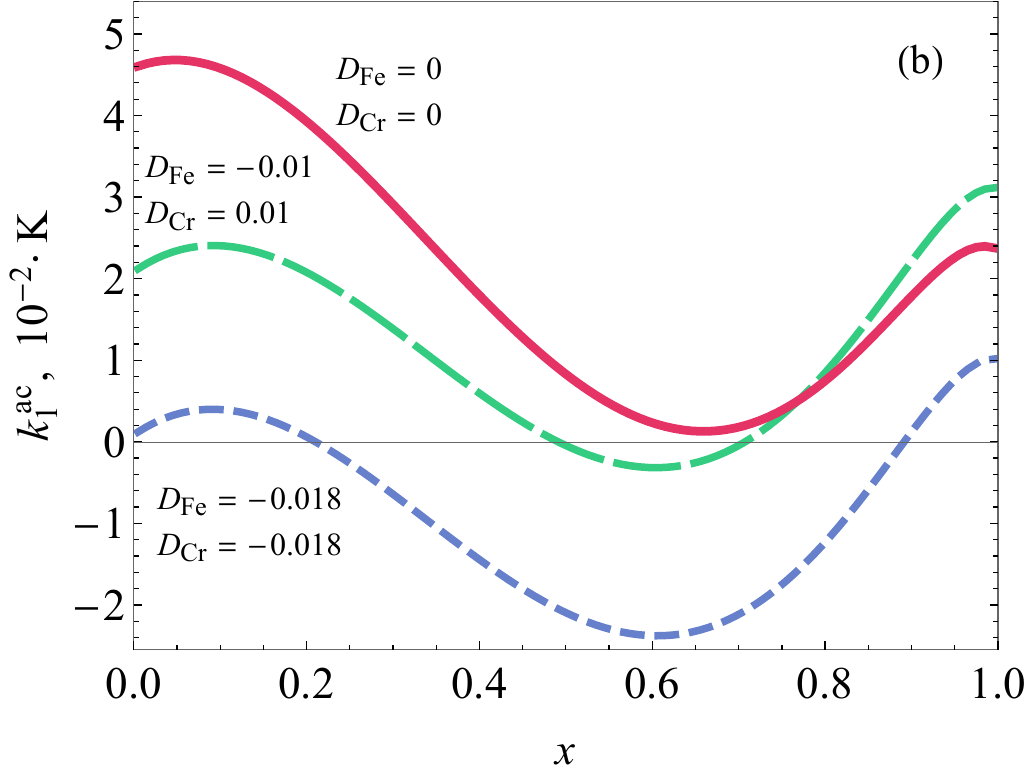}
\caption{
Concentration dependence of the effective anisotropy constant in the $ac$ plane: a) contribution of the DM interaction at different values of the parameter $d_{FeCr}$; b) the result of taking into account single-ion anisotropy at $d_{FeCr}=-2.5$\,K
\label{k1T0}}
\end{figure}

\section{The nature of spin reorientation in weak ferrimagnets of the YF\lowercase{e}$_{1-x}$C\lowercase{r}$_x$O$_3$ type}

Unlike YFeO$_3$ and YCrO$_3$, which are weak ferromagnets with a basic magnetic structure of the $G_xF_z$ type below $T_N$, weak ferrimagnetic orthoferrites-orthochromites YFe$_{1-x}$Cr$_x$O$ _3$, according to magnetic measurement data, reveal complete or partial spin reorientation of the $G_xF_z$--\,$G_zF_x$ type in a wide substitution range\,\cite{WFIM-1}. This unexpected behavior, usually typical of orthoferrites with magnetic rare-earth ions (Er, Tm, Dy, ...)\,\cite{SR-2022,SR-2023}, is explained mainly by a strong decrease in the contribution of the DM interaction to the magnetic field. anisotropy.

\begin{figure*}[t]
\centering
\includegraphics[width=0.98\textwidth]{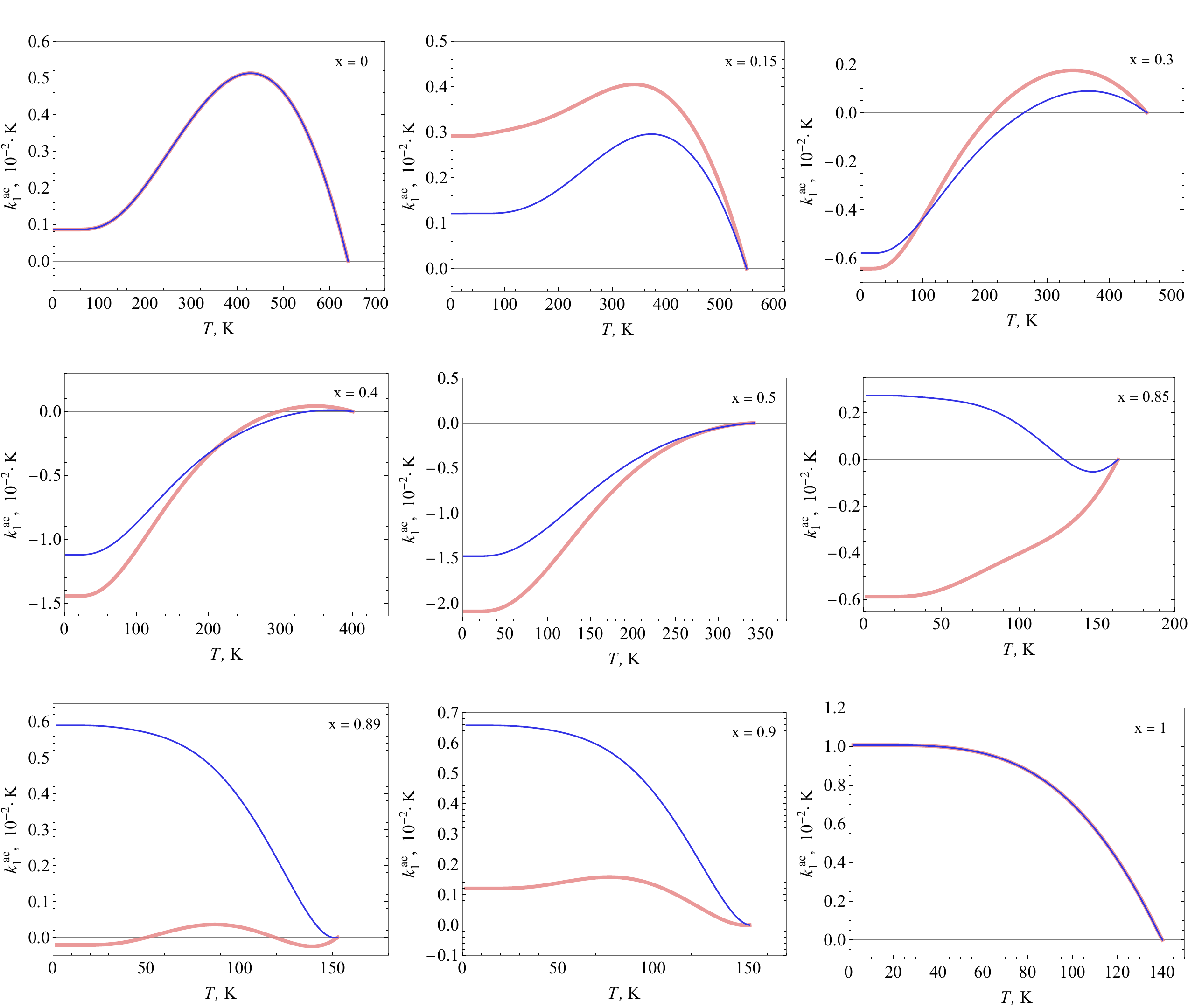}
\caption{
Calculated temperature dependencies effective constants  anisotropy second order $k_1(ac)$ for YFe$_{1-x}$Cr$_x$O$_3$ at $d_{FeCr}=-2.5$\,K, $D_{Fe} = D_{Cr} = -0.018$\,K and various concentrations $x$: red thick line -- MFA-I, blue thin line -- MFA-II
\label{k1}}
\end{figure*}

The contribution of competing antisymmetric exchange to the magnetic anisotropy of weak ferrimagnets has an unusual concentration dependence. Thus, if in pure orthoferrite YFeO$_3$ and orthochromite YCrO$_3$ antisymmetric exchange makes a decisive contribution to the stabilization of the magnetic configuration $\Gamma_4$, then in a weak ferrimagnet YFe$_{1-x}$Cr$_x$O$_3 $ it can cause a spin-reorientation transition $\Gamma_4$\,--\,$\Gamma_2$, characteristic of some orthoferrites RFeO$_3$ with magnetic rare earth ions (R\,=\,Nd, Sm, Tb, Ho, Er, Tm, Yb). Figure\,\ref{k1T0}a shows the concentration dependence of the contribution of the DM interaction to the first anisotropy constant for YFe$_{1-x}$Cr$_x$O$_3$ in the $ac$ plane for different values of the parameter $d_{FeCr}$, calculated within the framework of the simple MFA-I approximation in the low temperature limit. A characteristic feature of this dependence is the appearance of several extrema with a sharp decrease in the contribution in the region of intermediate concentrations near $x \sim 0.6-0.7$.

However, the appearance of spontaneous spin-reorientation transitions in weak ferrimagnets of the YFe$_{1-x}$Cr$_x$O$_3$ type with a non-magnetic R-ion is not only the result of competition between the Dzyaloshinsky-Moriya interactions Fe--Fe, Cr--Cr and Fe--Cr, but also comparable in magnitude, but ``opposite'' in sign, contributions from single-ion anisotropy of the Fe and Cr sublattices. For illustration, Fig.\,\ref{k1T0}b shows the result of competition between DM interaction and single-ion anisotropy in the $ac$ plane at $d_{FeCr}=-2.5$\,K and various single-ion anisotropy constants in the form $H_ {SIA}^{(2)}=D\,S_z^2$\,\cite{Taylor}. It is obvious that with an increase in the single-ion contribution, the region where the sign of the effective anisotropy constant changes in the $ac$ plane and the appearance of both ``flat'' angular phases of the $G_{xz}$ type and spatial ones of the $G_{xyz}$ type appears and expands.

Effective anisotropy constants in weak ferrimagnets have a specific temperature dependence\,\cite{Agafonov}. Figure\,\ref{k1} shows the calculated temperature dependences of the effective second-order anisotropy constant $k_1(ac)$ for YFe$_{1-x}$Cr$_x$O$_3$ at $d_{FeCr}= -2.5$\,K, $D_{Fe} = -0.018$\,K, $D_{Cr} = -0.018$\,K and various concentrations $x$. Let us note the significant difference in the results when using the MFA-I and MFA-II models for a number of concentrations.

\begin{figure}[h]
\centering
\includegraphics[width=0.48\textwidth]{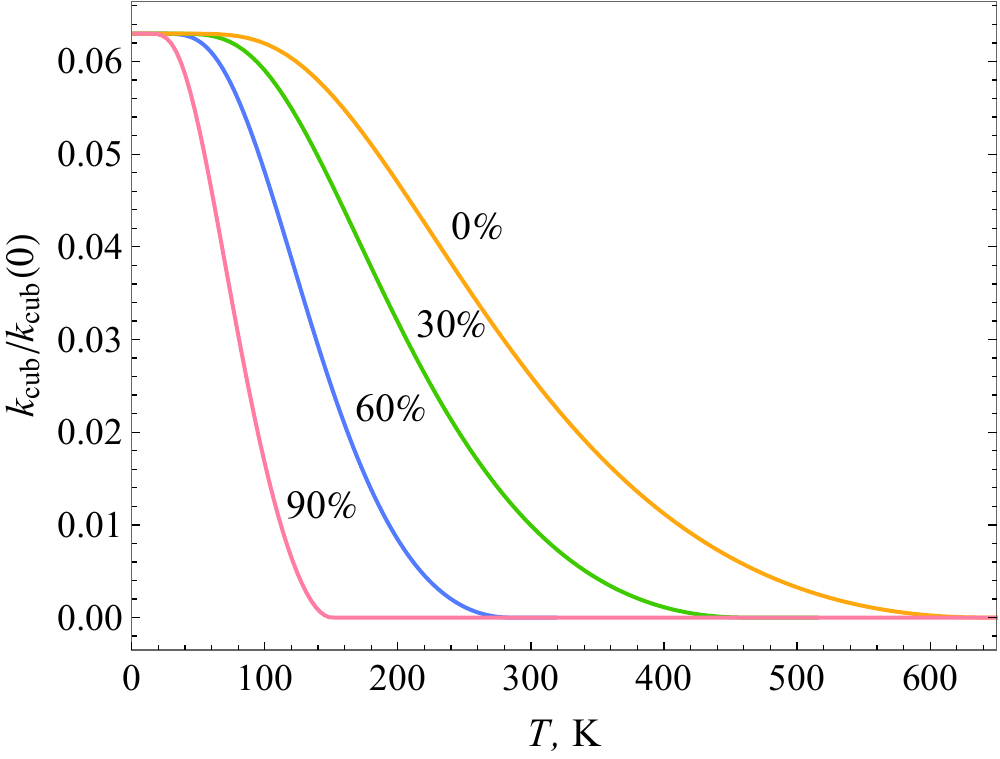}
\caption{
Temperature dependence of fourth-order single-ion anisotropy $k_{2} (ac)$ at concentrations $x = 0,\ 0.3,\ 0.6,\ 0.9$
\label{kcub}}
\end{figure}

The nature of spin-reorientation transitions significantly depends on the magnitude and sign of the fourth-order single-ion anisotropy constants\,\cite{SR-2022,SR-2023}. Fig.\,\ref{kcub} shows the temperature dependence of the fourth-order single-ion anisotropy constants $k_{cub}$ (\ref{kA3}) for a number of concentrations in YFe$_{1-x}$Cr$_x$O$ _3$.

In contrast to the yttrium system, lutetium orthoferrites-orthochromites LuFe$_{1-x}$Cr$_x$O$_3$ ($x = 0.0,\ 0.1,\ 0.2,\ 0.5,\ 0.6,\ 1.0$) retain the main magnetic structure of the $G_xF_z$ type without signs of a spontaneous spin-reorientation transition\,\cite{YLuFeCrO3}. As noted above, in YFeO$_3$ and LuFeO$3$ the contribution of single-ion anisotropy to the first anisotropy constant $k_1(ac)$ differs in sign, so that in YFeO$_3$ this contribution partially compensates for the contribution of the Dzyaloshinsky-Moriya interaction, and in LuFeO$ _3$ both contributions add up, resulting in a sharp increase in $k_1(ac)$\,\cite{thesis,Mukhin,JETP-1981} and greater stability of the $\Gamma_4$ configuration.

\section{Features of magnetization of weak ferrimagnets RF\lowercase{e}$_{1-x}$C\lowercase{r}$_x$O$_3$ with non-magnetic R-ion}

\subsection{Compensation points and negative magnetization}

One of the most interesting properties of weak ferrimagnets is the specific temperature and concentration dependences of magnetization, which are a consequence of the competition between Dzyaloshinsky Fe--Fe, Cr--Cr and Fe--Cr (Cr--Fe) interactions. Thus, back in 1977, it was predicted\,\cite{WFIM-1}, and in 1978, for the first time experimentally discovered\,\cite{WFIM-2}, the phenomenon of temperature compensation of the magnetic moment in a single-crystal sample of a weak ferrimagnet YFe$_{ 1-x}$Cr$_x$O$_3$ ($x = 0.38$) at $T_{comp} = 225$\,K. Much later, this effect (magnetization reversal, negative magnetization) was discovered in polycrystalline samples of yttrium orthoferrite-orthochromite at $x =0.5$ ($T_{comp} = 248$\,K\,\cite{Mao2011}, $T_{comp } \approx 230$\,K\,\cite{Dasari}, $T_{comp}=245$\,K\,\cite{Salazar2022}, $T_{comp}=175$\,K\,\cite {Yang}), at $x=0.4$ ($T_{comp}=170$\,K\,\cite{Dasari}), as well as lutetium orthoferrite-orthochromite at $x=0.5$ ($T_{comp} =224$\,K\,\cite{Pomiro}, $T_{comp}=230$\,K\,\cite{PRB-2018}, $T_{comp}=142$\,K\,\cite {Yang}).

It is interesting that the appearance of two points of concentration, and therefore temperature compensation, previously predicted by Kadomtseva and co-authors in weak ferrimagnets of the RFe$_{1-x}$Cr$_x$O$_3$ type with a non-magnetic R-ion (R = La, Y , Lu)\,\cite{WFIM-1} was experimentally confirmed in LaFe$_{1-x}$Cr$_x$O$_3$, where the magnetization inversion effect was observed both for compositions with $x=0.50$ and $ x=0.55$, and at concentration $x$ in the range $0.85-0.95$\,\cite{Bora}. The results of the work\,\cite{Dahmani2002} also indicate the detection of temperature compensation of magnetization in YFe$_{1-x}$Cr$_x$O$_3$ as at $x=0.50$ ($T_{comp} \approx 250 $\,K), and for $x=0.75$ ($T_{comp} \approx 50$\,K). The authors of the work\,\cite{Dasari} briefly mention the possible compensation of magnetization for YFe$_{1-x}$Cr$_x$O$_3$ in the composition range $x\sim0.8$.
 
Note that the compensation temperature $T_{comp}$, more precisely, the magnetization reversal temperature $T_{mr}$ significantly depends on the magnitude and method of switching on (ZFC, FC) the external field (see, for example, work\,\cite{Bora}), and for polycrystalline samples heat treatment can play a major role, which generally leads to both a scatter in the experimental values of $T_{mr}$ and ambiguity in quantitative theoretical analysis. It is interesting that even in the first experimental works attention was drawn to the ambiguity of data on magnetization at $T=77$\,K even for single-crystal samples with a nominal composition of YFe$_{0.5}$Cr$_{0.5}$O$_3$ and close Neel temperatures\,\cite{WFIM-1}.

Observation of the compensation point significantly depends on the magnitude of the external magnetic field (FC mode). Thus, the compensation temperature observed in LuFe$_{0.5}$Cr$_{0.5}$O$_3$ at $T_{comp}=224$\,K in a field of $100$\,Oe is suppressed in higher fields. In low fields, the total magnetic moment of a weak ferrimagnet in the FC mode, initially oriented in the direction of the applied field, is reoriented in the direction opposite to the applied magnetic field as the temperature decreases below $T_{comp}$. However, a sufficiently strong external magnetic field can hold the magnetic moment of the sample in the direction of the field\,\cite{Mao2011,Pomiro}.

By the way, due to the presence of concentration and temperature compensation points in weak ferrimagnets, the problem of the ``sign'' of the experimentally measured magnetization appears, ignoring which can lead to an erroneous interpretation of the experimental data.

\subsection{Competition of signs of Dzyaloshinsky vectors $d_{FeFe}$, $d_{CrCr}$ and $d_{FeCr}$ and features of concentration and temperature dependences of magnetization in YFe$_{1-x}$Cr$_x$O$ _3$}

We used the MFA-I and MFA-II methods outlined above to calculate the concentration and temperature dependences of the total magnetization of weak ferrimagnets and weak ferromagnetic moments of the sublattices $m_{Fe}, m_{Cr}$, or in other words, small components of the basis vector of ferromagnetism ${\bf F}$ (overt canting) for Fe- and Cr-sublattices in magnetic configurations $\Gamma_4$ and $\Gamma_2$. In a similar way, small components of the basis vectors ${\bf A}$ and ${\bf C}$ can be calculated (hidden canting) for Fe and Cr sublattices in magnetic configurations $\Gamma_{1,2,4}$.

\begin{figure}[h]
\centering
\includegraphics[width=0.48\textwidth]{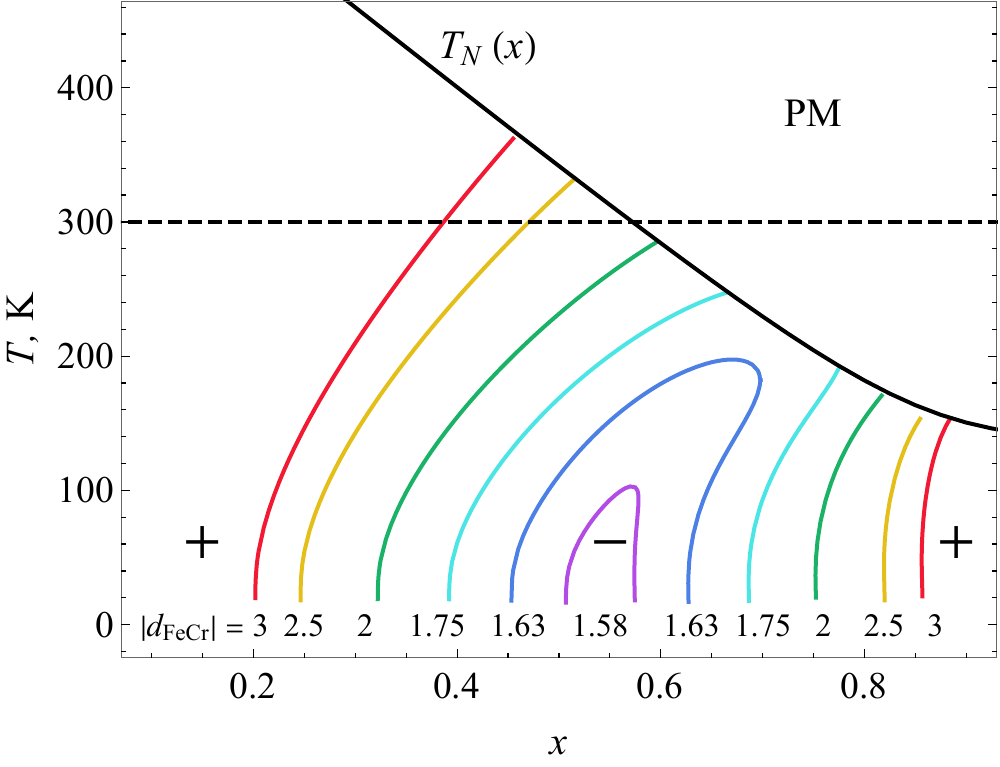}
\caption{
$T$--$x$ phase diagram of magnetization compensation regions of a weak ferrimagnet YFe$_{1-x}$Cr$_x$O$_3$ for different values of the parameter $d_{FeCr}$. The ``+'' sign indicates the region of ``positive'' magnetization. The dotted line highlights the room temperature $T=300$\,K
\label{TXcomp}}
\end{figure}

When solving the molecular field equations numerically, we assume that the parameters of both the isotropic superexchange interaction and the Dzyaloshinsky-Moriya interaction do not depend on the concentration $x$, so that $I_{FeFe}, I_{CrCr}, d_{FeFe}, d_{ CrCr}$ are close to their values in the parent compositions YFeO$_3$ and YCrO$_3$. A definite argument in favor of this assumption is the data from a neutron diffraction study of the YFe$_{0.5}$Cr$_{0.5}$O$_3$ system\,\cite{Nair,Yang}, indicating the similarity of the structural parameters in this mixed orthoferrite-orthochromite with data for the parent YFeO$_3$ and YCrO$_3$. Indeed, the values of the structure factor that determines the orientation of the Dzyaloshinsky vector, calculated from the data\,\cite{Yang} (Table\,\ref{rxr}), coincide with good accuracy with the corresponding data for YFeO$_3$ (see, for example, Table\,3 in the work\,\cite{M-2021}), which justifies the possibility of using constant, concentration-independent, values of Dzyaloshinsky vectors. In accordance with the predictions of microscopic theory, the Dzyaloshinsky vector ${\bf d}_{FeCr}$ has a comparable magnitude and opposite direction to the vectors ${\bf d}_{FeFe}$ and ${\bf d}_{CrCr}$ (see Table\,\ref{tablesign}\,\cite{1977,JMMM-2016,CM-2019,JETP-2021}).
For the exchange integral $I_{FeCr}=I_{FeCr}$ we use the data from the work\,\cite{Ovanesyan}: $I_{FeCr}=13.4 \pm 0.4$. The scalar factors $d_{FeFe}(\theta)=2.0$\,K and $d_{CrCr}(\theta)=1.7$\,K were determined from low-temperature magnetization data of the parent compositions.

The calculation results showed that when $d_{FeCr}$ differs in sign from $d_{FeFe}$ and $d_{CrCr}$, the magnetization of YFe$_{1-x}$Cr$_x$O$_3$ drops sharply with the deviation from the parent compositions, but at $|d_{FeCr}|\geq |d_{FeCr}^{(cr)}|$, where $d_{FeCr}^{(cr)} \approx -1.55$\,K, on the $T$--$x$ phase diagram, a region of negative magnetization with two compensation points appears and grows with increasing $|d_{FeCr}|$. Fig.\,\ref{TXcomp} shows the $T$--$x$ phase diagram of the weak ferrimagnet YFe$_{1-x}$Cr$_x$O$_3$ (MFA-II), where the $T_N(x)$ curve limits region of magnetic ordering, and thin lines indicate lines of compensation points, that is, changes in the sign of magnetization at different values of the parameter $d_{FeCr}$.

\begin{figure*}[t]
\centering
\includegraphics[width=0.98\textwidth]{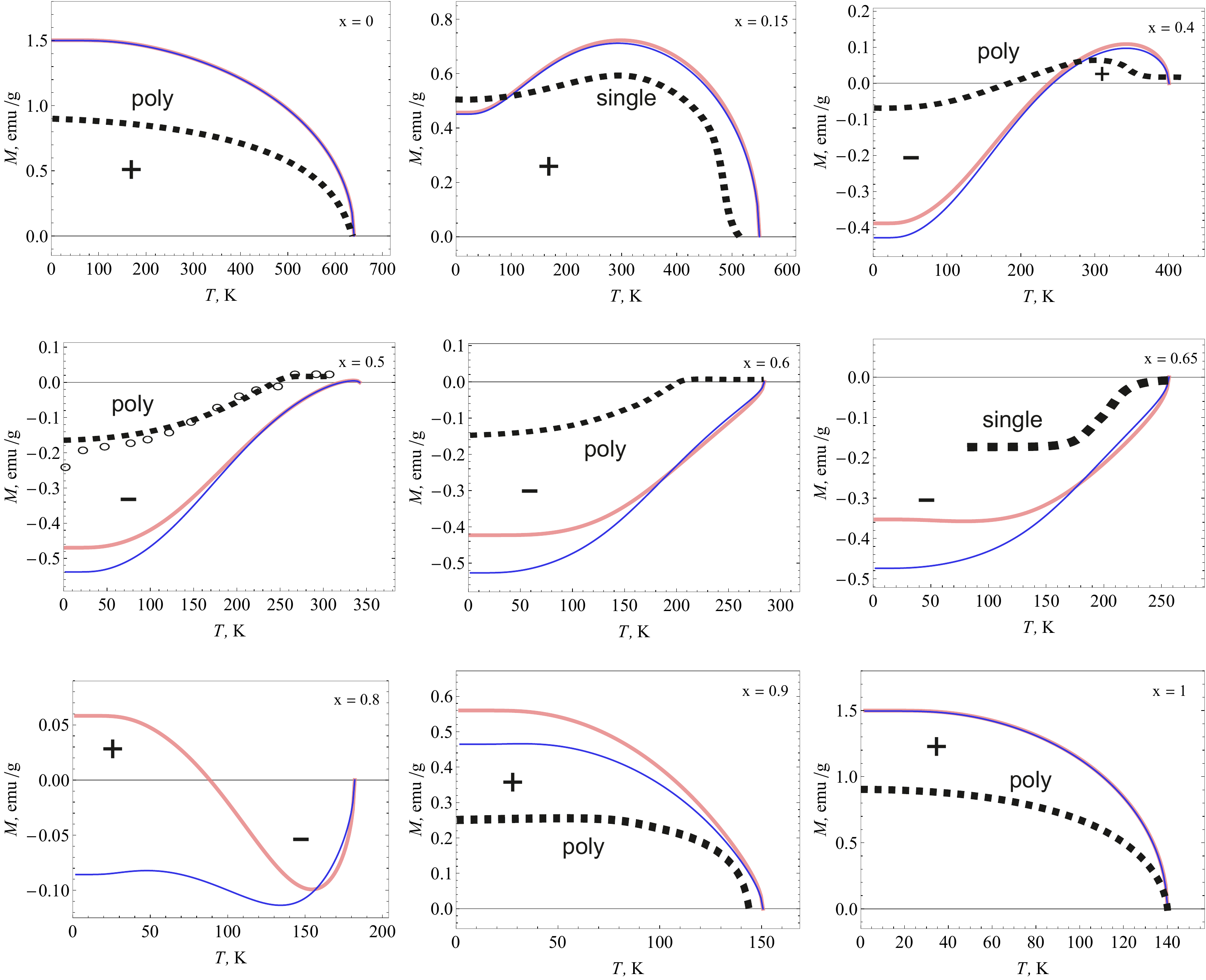}
\caption{
Temperature dependence of magnetization in a model weak ferrimagnet YFe$_{1-x}$Cr$_x$O$_3$ at $x=0.0,\ 0.15,\ 0.4,\ 0.5,\ 0.6,\ 0.65,\ 0.8,\ 0.9 ,\ 1.0$: dotted lines -- experimental data, solid lines -- theory (thick red -- MFA-I, thin blue -- MFA-II)
\label{M(T)}}
\end{figure*}

Figure \,\ref{M(T)} shows the results of calculating the temperature dependences of the magnetization of a weak ferrimagnet YFe$_{1-x}$Cr$_x$O$_3$ at certain concentration values from $x=0$ to $ x=1$ within the framework of the MFA-I and MFA-II models and under the assumption that the magnetic configuration $\Gamma_4(G_x)$ is conserved throughout the concentration range. The scalar factor $d_{FeCr}(\theta) = d_{CrFe}(\theta) = -2.5$\,K was chosen from the condition that the compensation point for the composition $x=0.38$ appears near $T\approx 225$\, K. Naturally, both model approaches give identical or close results for compositions with $x\leq0.5$ and $x\geq0.9$, but the deviation increases as we approach compositions with $x\sim 0.7$. As one would expect, the theory predicts two regions of temperature compensation -- near compositions with $x\sim 0.4$ (wide ``high-temperature'' region) and $x\sim 0.8$ (narrow ``low-temperature'' region). For a composition with $x\sim 0.45$, compensation is expected near room temperature.

Unfortunately, there are no data in the literature on regular measurements of the temperature dependences of the magnetization of single-crystal samples of weak ferrimagnets of the YFe$_{1-x}$Cr$_x$O$_3$ type in a wide range of concentrations, with the exception of compositions with $x=0.15$, 0.38, 0.5, 0.65, for which some information is available\,\cite{WFIM-1,WFIM-2}. Comparison of the calculated data with the measurement data of the total magnetization for the composition $x=0.15$ (see Fig.\,\ref{M(T)}) shows quite satisfactory agreement, despite the presence of spin reorientation $G_x\leftrightarrow G_z$ in the range of $250-400$\,K. The saturation magnetization of single-crystal samples with $x=0.38$, 0.5, 0.65 is significantly, two to three times, less than the predictions of the molecular field theory, which, in light of the experimentally discovered transitions for these compositions with a change in the orientation of the weak ferrimagnetic moment in the $ac$ plane \,\cite{WFIM-1} indicates the possible implementation of the spatial orientation of the antiferromagnetism vector, that is, the $G_{xyz}$ configuration.

Magnetic measurements on polycrystalline samples, especially under conditions of spin-reorientation states, extremely small values of magnetization near the compensation region and negative magnetization, cannot provide reliable values of Neel temperatures, compensation temperatures and absolute values of magnetization. Nevertheless, the data of calculations within the framework of molecular field theory with only one fixed parameter $d_{FeCr}$, presented in Fig.\,\ref{M(T)} provide a semi-quantitative explanation of the experimental data of the work\,\cite{Dasari}, obtained on polycrystals of the weak ferrimagnet YFe$_{1-x}$Cr$_x$O$_3$ in a wide range of concentrations $x=0 - 1$.

\begin{figure*}[t]
\centering
\includegraphics[width=0.3\textwidth]{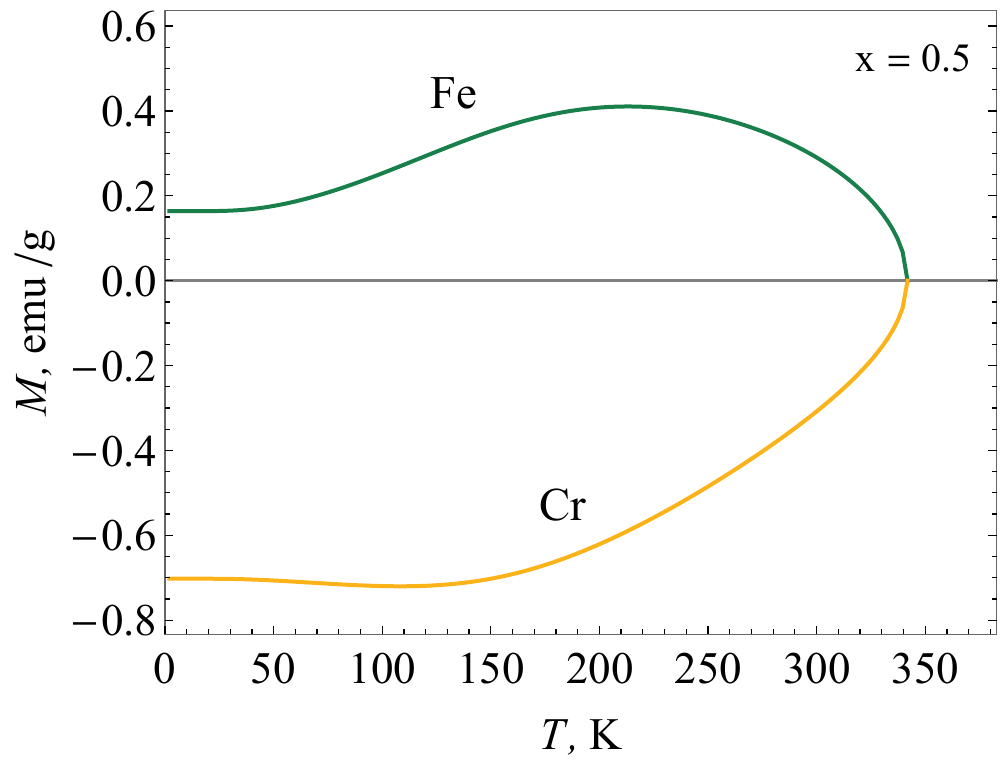}
\includegraphics[width=0.3\textwidth]{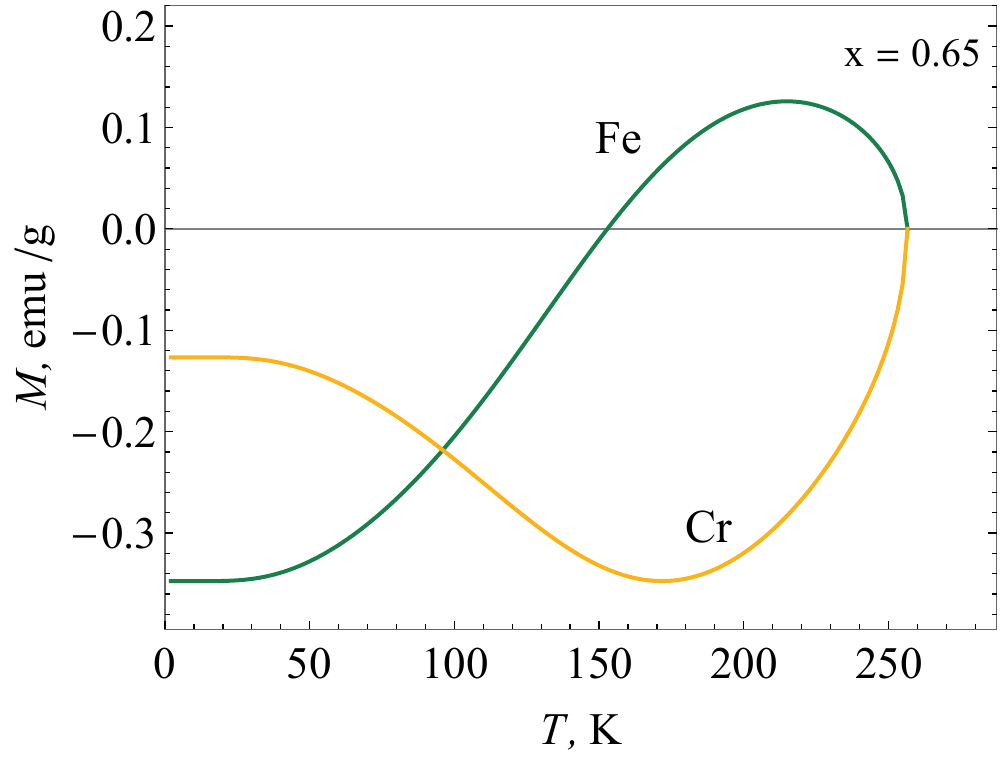}
\includegraphics[width=0.3\textwidth]{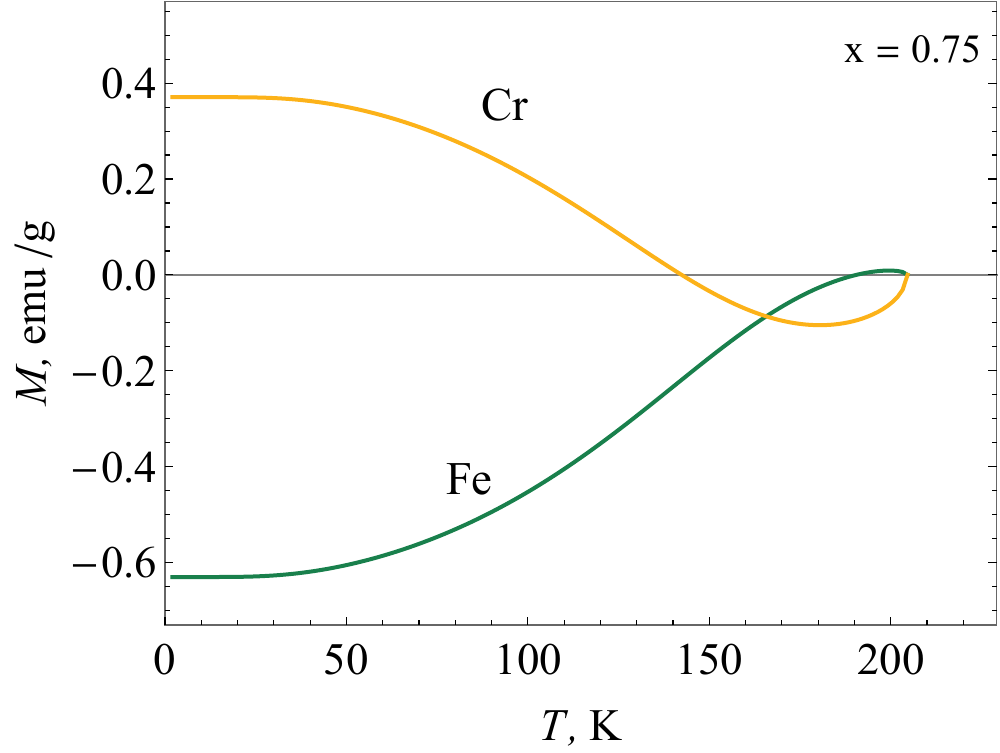}
\caption{
Temperature dependences of partial magnetizations $m_{Fe}$ and $m_{Cr}$ in the weak ferrimagnet YFe$_{1-x}$Cr$_x$O$_3$ for compositions $x =0.5,\ 0.65,\ 0.75$
\label{Mpartial}}
\end{figure*}

However, the model approaches and formulas used to describe the temperature dependences of the magnetization of weak ferrimagnets of the YFe$_{1-x}$Cr$_x$O$_3$ type in the works\,\cite{Dasari,Billoni} (see formulas (6) and (12) in \,\cite {Billoni}) do not follow from any rigorous MFA approach, but are merely the result of a ``plausible'' phenomenology using a number of, sometimes hidden, parameters. Thus, the apparently impressive agreement between theory and experiment in the work\,\cite{Dasari} (see Figures 5-7 of this work) is based on a fit with several parameters, including, as noted above, the unphysically large value of the exchange integral $I_{FeCr }$, as well as a non-physically strong concentration dependence of the $d_{FeCr}$ parameter of the Dzyaloshinsky-Moriya interaction, with a sharp drop from $d_{FeCr}=-1.3$\,K at $x=0.1$ to almost zero at $x= 0.9$. There are a number of inconsistencies in the work, for example, the ``theoretical'' Neel temperatures for the composition with $x=0.3$ in Fig.\,2 and in Fig.\,5 differ by 50\,K. But moreover, the authors incorrectly attributed a positive sign to the magnetization of compositions with $x=0.6$ and $x=0.7$, and were unable to explain the observed compensation of magnetization at $x=0.8$ and the ``return'' to the positive sign of magnetization at $x= 0.9$. The authors do not consider the effects of spin reorientation and the ``symmetrical'' contribution of single-ion anisotropy to magnetization at all. In general, the authors' statement ``Thus, we are able to obtain quantitative agreement between theory and agreement for the whole range of doping and temperature with a very simple, consistent and transparent approach'' is more than controversial.
    
Let us emphasize once again that the features of the concentration and temperature dependences of the magnetization of weak ferrimagnets of the YFe$_{1-x}$Cr$_x$O$_3$ type, in particular the compensation phenomenon, are determined primarily by the competition of the Dzyaloshinsky-Moriya Fe--Fe interactions, Cr--Cr and Fe--Cr, although the contribution of single-ion spin anisotropy can have a noticeable effect on both the offset of the compensation temperature and the magnitude and nature of the temperature dependence of magnetization near $T_{comp}$.

It is interesting that not only the total magnetization, but also the partial magnetization of the Fe and Cr sublattices, $m_{Fe}$ and $m_{Cr}$ have unusual concentration and temperature dependences. For illustration, Fig.\,\ref{Mpartial} shows the temperature dependences of $m_{Fe}$ and $m_{Cr}$ for compositions $x = 0.5$, 0.65, 0.75 calculated in the MFA-II approximation. If for the composition $x = 0.5$ (as well as for $x<0.5$) ferrimagnetic ordering of weak ferromagnetic moments of the Fe-Cr sublattices is realized, then for compositions with $x =0.65$, 0.75 we discover the phenomenon of temperature compensation for partial contributions.

\section{Conclusion}

Weak ferrimagnets with competing Dzyaloshinsky-Moriya interactions of the RFe$_{1-x}$Cr$_x$O$_3$ type with magnetic and non-magnetic R-ions represent a new fairly wide class of promising magnetic materials with a number of unique magnetic, magnetocaloric, and magnetoelectric properties.

In this paper, we present a brief critical review of the 50-year history of experimental and theoretical studies of the magnetic properties of weak ferrimagnets of the YFe$_{1-x}$Cr$_x$O$_3$ type. The spin Hamiltonian of the system is considered taking into account the main isotropic and anisotropic interactions. Within the framework of the molecular field approximation, calculations were made of the Neel temperatures, the average magnetic moments of $3d$ ions, total and partial magnetizations, and effective anisotropy constants. It is shown that in the model system YFe$_{1-x}$Cr$_x$O$_3$ there are two regions of negative magnetization $0.25 \leq x\leq 0.5$ and $x\approx 0.8$ with corresponding compensation points, reaching room temperature at $x\approx 0.45$. The phenomenon of spin reorientation observed for single-crystal samples in a wide range of concentrations is explained by a sharp decrease in the contribution of antisymmetric exchange to magnetic anisotropy with increasing deviation from the parent compositions and competition between the contributions of single-ion anisotropy of Fe and Cr ions. It has been suggested that the spatial orientation of the antiferromagnetism vector and the $G_{xyz}$ configuration are the reason for the small value of saturation magnetization observed experimentally for compositions inside or near the region of negative magnetization.

Understanding the mechanisms of formation of the peculiarities of concentration and temperature dependences of magnetization, primarily the phenomenon of compensation and negative magnetization, the phenomenon of spin reorientation, is the basis for a scientifically based approach to controlling these properties, predicting and synthesizing new promising materials.




\begin{acknowledgments}
This study was supported by the Ministry of Science and Higher Education of the Russian Federation, project FEUZ-2023-0017
\end{acknowledgments}

\appendix
\section{Magnetic moment equations in the MFA-II model}\label{MFAIIparams}

Phase $\Gamma_4 (G_x, A_y, F_z)$:
\begin{gather*}
m^z_{Fe} (\alpha_{Fe}^{+} + 1) + m^z_{Cr} \beta_{Fe}^{+} = -\gamma_{Fe}^F\,; \\
m^z_{Fe} \alpha_{Cr}^{+} + m^z_{Cr} (\beta_{Cr}^{+} + 1) = -\gamma_{Cr}^F \,; \\
m^y_{Fe} (\alpha_{Fe}^{-} + 1) + m^y_{Cr} \beta_{Fe}^{-} = \gamma_{Fe}^A\,; \\
m^y_{Fe} \alpha_{Cr}^{-} + m^y_{Cr} (\beta_{Cr}^{-} + 1) = \gamma_{Cr}^A \,;
\end{gather*}
phase $\Gamma_2 (F_x, C_y, G_z)$:
\begin{gather*}
m^x_{Fe} (\alpha_{Fe}^{+} + 1) + m^x_{Cr} \beta_{Fe}^{+} = \gamma_{Fe}^F\,; \\
m^x_{Fe} \alpha_{Cr}^{+} + m^x_{Cr} (\beta_{Cr}^{+} + 1) = \gamma_{Cr}^F\,;\ \\
m^y_{Fe} (\alpha_{Fe}^{-} - 1) + m^y_{Cr} \beta_{Fe}^{-} = \gamma_{Fe}^C\,; \\
m^y_{Fe} \alpha_{Cr}^{-} + m^y_{Cr} (\beta_{Cr}^{-} - 1) = \gamma_{Cr}^C \,;\
\end{gather*}
phase $\Gamma_1 (A_x, G_y, C_z)$:
\begin{gather*}
m^x_{Fe} (\alpha_{Fe}^{-} + 1) + m^x_{Cr} \beta_{Fe}^{-} = -\gamma_{Fe}^A\,; \\
m^x_{Fe} \alpha_{Cr}^{-} + m^x_{Cr} (\beta_{Cr}^{-} + 1) = -\gamma_{Cr}^A \,;\ \\
m^z_{Fe} (\alpha_{Fe}^{+} - 1) + m^z_{Cr} \beta_{Fe}^{+} = -\gamma_{Fe}^C\,; \\
m^z_{Fe} \alpha_{Cr}^{+} + m^z_{Cr} (\beta_{Cr}^{+} - 1) = -\gamma_{Cr}^C \,;\
\end{gather*}
where
\begin{gather*}
\alpha_{Fe}^{\pm} = \sum_{klr} {( k + l \pm r )\,I_{FeFe}}\, P^{klr}_{Fe}; \\
\alpha_{Cr}^{\pm} = \sum_{klr} {( k + l \pm r )\,I_{FeCr}}\, P^{klr}_{Cr}\,;\ \\
\beta_{Fe}^{\pm} = \sum_{klr} {(2u - k - l \pm u \mp r)\,I_{FeCr}}\, P^{klr}_{Fe}; \\
\beta_{Cr}^{\pm} = \sum_{klr} {(2u - k - l \pm u \mp r)\,I_{CrCr}}\, P^{klr}_{Cr}\, ;\
\end{gather*}
\begin{gather*}
\gamma_{Fe}^F = \sum_{klr} \Big\{ \big[ (k + l)\,d_{FeFe}^{\, y_4} + r\,d_{Fe}^{\, y_2 } \big] \langle S_{Fe} \rangle + \\
+ \big[ (2u - k - l)\,d_{FeCr}^{\, y_4} + (u - r)\,d_{FeCr}^{\, y_2} \big] \langle S_{Cr} \rangle \Big\} P^{klr}_{Fe};\ \\
\gamma_{Cr}^F = \sum_{klr} \Big\{ \big[ (k + l)\,d_{FeCr}^{\, y_4} + r\,d_{FeCr}^{\, y_2 } \big] \langle S_{Fe} \rangle + \\
+ \big[ (2u - k - l)\,d_{CrCr}^{\, y_4} + (u - r)\,d_{CrCr}^{\, y_2} \big] \langle S_{Cr} \rangle \Big\} P^{klr}_{Cr};\ \\
\gamma_{Fe}^A = \sum_{klr} \Big\{ (k + l)\,d_{FeFe}^{\, z_4} \,\langle S_{Fe} \rangle + \\
+ (2u - k - l)\,d_{FeCr}^{\, z_4} \,\langle S_{Cr} \rangle \Big\} P^{klr}_{Fe};\ \\
\gamma_{Cr}^A = \sum_{klr} \Big\{ (k + l)\,d_{FeCr}^{\, z_4}\,\langle S_{Fe} \rangle + \\
+ (2u - k - l)\,d_{CrCr}^{\, z_4} \,\langle S_{Cr} \rangle \Big\} P^{klr}_{Cr};\ 
\end{gather*}
\begin{gather*}
\gamma_{Fe}^C = \sum_{klr} \Big\{ \big[ (k - l)\,|d_{FeFe}^{\, x_4}| + r\,d_{FeFe}^{\, x_2} \big] \langle S_{Fe} \rangle + \\
+ \big[ (l - k)\,|d_{FeCr}^{\, x_4}| + (u - r)\,d_{FeCr}^{\, x_2} \big] \langle S_{Cr} \rangle \Big\} P^{klr}_{Fe};\ \\
\gamma_{Cr}^C = \sum_{klr} \Big\{ \big[ (k - l)\,|d_{FeCr}^{\, x_4}| + r\,d_{FeCr}^{\, x_2} \big] \langle S_{Fe} \rangle + \\
+ \big[ (l - k)\,|d_{CrCr}^{\, x_4}| + (u - r)\,d_{CrCr}^{\, x_2} \big] \langle S_{Cr} \rangle \Big\} P^{klr}_{Cr};\ \\
P^{klr}_{Fe} = \frac{p^{klr}}{h^{klr}_{Fe}}\,S_{Fe}\,B_{S_{Fe}} \left( \frac {S_{Fe} h^{klr}_{Fe}}{k_B T} \right); \\
P^{klr}_{Cr} = \frac{p^{klr}}{h^{klr}_{Cr}}\,S_{Cr}\,B_{S_{Cr}} \left( \frac {S_{Cr} h^{klr}_{Cr}}{k_B T} \right) ;\ \\
h^{klr}_{Fe} = (k + l + r)I_{FeFe} \langle S_{Fe} \rangle + (z - k - l - r)I_{FeCr} \langle S_{Cr} \rangle ;\ \\
h^{klr}_{Cr} = (k + l + r)I_{FeCr} \langle S_{Fe} \rangle + (z - k - l - r)I_{CrCr} \langle S_{Cr} \rangle ;\ \\
\langle S_{Fe} \rangle = \sum_{klr} p^{klr}\,S_{Fe}\, B_{S_{Fe}} \left( \frac{S_{Fe} h^{klr}_ {Fe}}{k_B T} \right); \\
\langle S_{Cr} \rangle = \sum_{klr} p^{klr}\,S_{Cr}\, B_{S_{Cr}} \left( \frac{S_{Cr} h^{klr}_ {Cr}}{k_B T} \right) .
\end{gather*}
Additional index at Dzyaloshinsky parameters  indicates on number sublattices, for example $d^{\, y_2}_{FeFe} = d_{FeFe}\, [{\bf r}_2 \times {\bf r}_1]^y$ and $d^{\, z_4 }_{FeCr} = d_{FeCr}\, [{\bf r}_4 \times {\bf r}_1]^z$.

\newpage

\end{document}